\documentclass{elsart}
\usepackage{epsfig}
\usepackage{amssymb}

\def\be{\begin{eqnarray}}
\def\ee{\end{eqnarray}}
\def\lsim{\mathrel{\rlap{\lower3pt\hbox{\hskip1pt$\sim$}}
     \raise1pt\hbox{$<$}}} 
\def\gsim{\mathrel{\rlap{\lower3pt\hbox{\hskip1pt$\sim$}}
     \raise1pt\hbox{$>$}}} 

\def\cal{\it}

\begin{document}

\runauthor{Brown, Lee, \& Rho}

\begin{frontmatter}
\title{Chemical Equilibration\\ in Relativistic Heavy Ion Collisions}

\author[suny]{Gerald E. Brown,}
\author[pnu]{Chang-Hwan Lee}
\author[saclay]{and Mannque Rho}

\address[suny]{Department of Physics and Astronomy,\\
               State University of New York, Stony Brook, NY 11794, USA \\
(\small E-mail: Ellen.Popenoe@sunysb.edu)}

\address[pnu]{Department of Physics, \\
and Nuclear physics \& Radiation technology Institute (NuRI),\\
Pusan National University,
              Pusan 609-735, Korea\\ (E-mail: clee@pusan.ac.kr) }
\address[saclay]{Service de Physique Th\'eorique, CEA Saclay, 91191 Gif-sur-Yvette
c\'edex, France\\
\& Department of Physics, Hanyang University, Seoul 133-791, Korea\\
(E-mail: rho@spht.saclay.cea.fr)}

\renewcommand{\thefootnote}{\fnsymbol{footnote}}
\setcounter{footnote}{0}

\begin{abstract}

In the hadronic sector of relativistic heavy ion physics, the
$\rho\leftrightarrows 2\pi$ reaction is the strongest one, strong
enough to equilibrate the $\rho$ with the pions throughout the
region from chemical freezeout to thermal freezeout when
free-particle interactions (with no medium-dependent effects) are
employed. Above the chiral restoration temperature, only $\rho$'s
and $\pi$'s are present, in that the chirally restored $A_1$ is
equivalent to the $\rho$ and the mesons have an SU(4) symmetry,
with no dependence on isospin and negligible dependence on spin.
In the same sense the $\sigma$ and $\pi$ are ``equivalent"
scalars. Thus the chirally restored $\rho\leftrightarrows 2\pi$
exhaust the interspecies transitions. We evaluate this reaction at
$T_c$ and find it to be much larger than below $T_c$, certainly
strong enough to equilibrate the chirally restored mesons just
above $T_c$. When emitted just below $T_c$ the mesons remain in
the $T_c+\epsilon$ freezeout distribution, 
at least in the chiral limit because of the
Harada-Yamawaki ``vector manifestation" that requires that mesonic
coupling constants go to zero (in the chiral limit) as $T$ goes to
$T_c$ from below. Our estimates in the chiral limit give
deviations in some particle ratios from the standard scenario (of
equilibrium in the hadronic sector just below $T_c$) of about
double those indicated experimentally.
This may be due to the neglect of explicit chiral symmetry
breaking in our estimates.
We also show that the
instanton molecules present above $T_c$ are the giant multipole
vibrations found by Asakawa, Hatsuda and Nakahara and of Wetzorke
et al. in lattice gauge calculations. Thus, the matter formed by
RHIC can equivalently be called: chirally restored mesons,
instanton molecules, or giant collective vibrations. It is a
strongly interacting liquid.

\end{abstract}

\end{frontmatter}

\renewcommand{\thefootnote}{\arabic{footnote}}
\setcounter{footnote}{0}
\section{Introduction\label{intro}}

One of the recurring results from relativistic heavy ion collisions,
emphasized by Braun-Munzinger and Stachel and collaborators,
found first in AGS and then at CERN energies and most recently at
RHIC \cite{BMRS01} has been the high degree of chemical equilibration
of the hadronic products of the collisions.
All of them, with exceptions we shall discuss,
freeze out in chemical composition at the same temperature,
essentially the temperature for chiral restoration $T_c$.

Many attempts to explain this equilibration in the accepted
scenario of the quark gluon plasma(QGP):
increased equilibration because of the higher
number of degrees of freedom in the QGP, or elsewhere, in the greater
number of degrees of the high temperature hadron gas because of dropping
masses, were ultimately defeated because of the small coupling constants
in these phases. In the QGP the perturbative coupling constants were
simply too small to bring about equilibration. Efforts in the hadronic
sector were defeated by the RG results of Harada and Yamawaki
\cite{HY:PR} that hadronic coupling constants go
to zero at $T_c$ in the chiral limit.

Recently Brown et al.\cite{BLRS} have shown that a second-order
phase transition can be constructed for RHIC physics (ignoring the
small baryon number in central collisions) by making the chirally
restored $\pi$ and $\sigma$ massless just above $T_c$. In this way
there is a continuity with the massless $\pi$ and $\sigma$, in the
chiral limit, just below $T_c$. Quark masses~\footnote{The ``quark
mass" we refer to in this paper is the chirally invariant thermal
mass, not the current quark mass that breaks chiral symmetry.},
measured in quenched lattice gauge studies \cite{petreczky02} are
$\gsim 1$ GeV both at $\frac 32 T_c$ and $3 T_c$. We shall assume
them to have this value\footnote{As noted in BLRS\cite{BLRS} with
our dynamic confinement the mass measured by the Polyakov loop is
never reached.} right down to $T_c$. In BLRS\cite{BLRS} at $T_c$,
the color Coulomb interaction brought the $\pi$- and
$\sigma$-masses down from 2 GeV to 1.5 GeV, and then the 4-point
instanton molecule interaction brought them the rest of the way to
zero.

The four-point interaction\footnote{The 4-point interaction is
modelled as a $\delta$-function with constant coefficient;
therefore, trivially factorizable.} acted as a driving force in an
RPA, with all bound $\bar q q$ states at unperturbed energy 1.5
GeV in a Furry representation (Coulomb eigenstates); i.e.,
quarkonium. Therefore Brown's \cite{brown67} degenerate schematic
model, in which eigenfunctions, etc., have simple analytic forms,
could be used.

It was pointed out that the resulting collective excitations, the giant
multipole states, were seen in the Asakawa et al.\cite{asakawa03}
LGS, also by Wetzorke et al.\cite{wetzorke}.

An important new development was the argument by Braun-Munzinger et al.
\cite{BMSW03}
that hadron multiplicities in central high-energy nucleus-nucleus
collisions are established essentially at
the phase boundary between
chirally broken and chirally restored matter. This must result
from multiparticle collisions, which are particularly strong just above
$T_c$.

In this note we develop the schematic model to include nonlinear
couplings of the giant vibrations just above $T_c$. We study, in
particular, the $\rho\rightarrow 2\pi$ reaction in the chirally
restored region, both the $\rho$ and $\pi$ being giant resonances.
This way of looking at the $\rho$ and $\pi$ was developed long ago
as the loop sum in the Nambu-Jona-Lasinio approach in the chirally
broken region of hadrons, but it is much simpler here because of
the degeneracy in Coulomb-bound $\bar q q$ states above $T_c$. We
note that the chirally restored $\pi$'s and $\sigma$'s are the
same entities, also the chirally restored vectors and axial
vectors. Therefore, we need consider only the interaction $\rho
\leftrightarrows 2\pi$ to determine the nonlinearity of the
vibrations, this covering all possibilities.


\section{The Free $\rho$ and $\pi$ as Giant Resonances; The Strong
$\rho\rightarrow 2\pi$ Transition as a Nonlinearity in the Vibrations}
\label{sec2}

Many research workers are not used to thinking of the $\rho$-meson
as a giant dipole resonance, but that will be a convenient place
to begin. The point is that the $\rho$, all by itself, is a
many-body problem in the Nambu-Jona-Lasinio language. It is a
giant dipole vibration of the negative energy sea, a sum of
quark-antiquark pieces.

\begin{figure}
\centerline{\epsfig{file=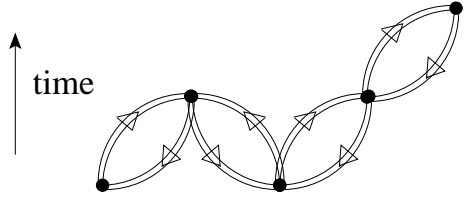,height=1.0in}}
\caption{
The sum of bubbles in NJL, which can go either forward or backward
in time, the latter representing ground state correlations. The solid
dots represent the vertex function, $\gamma_\mu$ for the
$\rho$, $\gamma_5$ for the $\pi$, multiplied by a $\delta$-function
and coupling constant.
The double lines indicate that quark and antiquark, or quark
and quark holes, are Coulomb bound states; i.e., are in the
Furry representation, which we will go over to above $T_c$.
}
\label{fig1}
\end{figure}

In NJL at zero density and temperature the negative energy states are filled
up to a cutoff $\Lambda\sim 700$ MeV. In the modern way of looking
at matters, this $\Lambda$ is the Wilsonian matching scale for
constituent quarks\cite{brown02}. The $\rho$ or $\pi$ are then
obtained as a sum of loop diagrams, the vertex function $\Gamma$ equal
to $\gamma_\mu$ or $\gamma_5$ for $\rho$ and $\pi$, respectively,
as in random phase approximation, as shown in Fig.~\ref{fig1},
except that in the chirally broken sector the Coulomb interaction
is unimportant.
The backward going lines represent holes in the negative energy sea,
the forward-going ones, particles.

A good complete review of these matters is given by Vogel \& Weise
\cite{vogl91}.

Now the width for the free $\rho$ to decay into two pions is large,
$\sim 150$ MeV, meaning that the lifetime of the $\rho$ is
$\tau\sim 4/3$ fm/c, still long enough that the $\rho$ can be distinguished
as a real particle. The point we wish to make is that the
$\rho\rightarrow 2\pi$ transition in medium
can be considered to be a nonlinearity,
a giant vector vibration changing into two giant pseudoscalar ones.
This nonlinear coupling will damp the $\rho$ vibrations, mixing the
energy between $\rho$ and $\pi$ ones, essentially powering the
many-body interactions which give the collectivity to these
vibrations.

Note that the $\rho\rightarrow 2\pi$ transition is already a
strong one in the broken symmetry sector. Also the $A_1\rightarrow
\rho +\pi$ has an even larger width, and can be considered the
same way as a collective excitation. However, the enhancement due
to collective effects, although large, is not nearly as large as
it would be if the unperturbed energy of all of the
particle-antiparticle loops were the same, as we shall find to be
the case in the chirally restored sector.

\section{Coulomb Bound States Above $T_c$: The Furry Representation}
\label{sec3}

Shuryak \& Zahed \cite{shuryak2003} introduced Coulomb bound states of
quark and antiquark, or of quark-particle and quark-hole for the
region of initial RHIC temperatures, suggesting that the breakup of
these molecules would help with the observed early equilibration.
Below $T\sim 2 T_c$ the $\bar q q$ states would be bound,
quite strongly as $T$ moves down towards $T_c$. The color Coulomb
interactions can be summed to all orders by going over to
quarkonium which has quark and antiquark in Coulomb bound states.
This is
the so called Furry representation, well known in atomic
physics, and shown by the double lines in Fig.~\ref{fig1}.
This gives us a convenient starting point.

\begin{figure}
\centerline{\epsfig{file=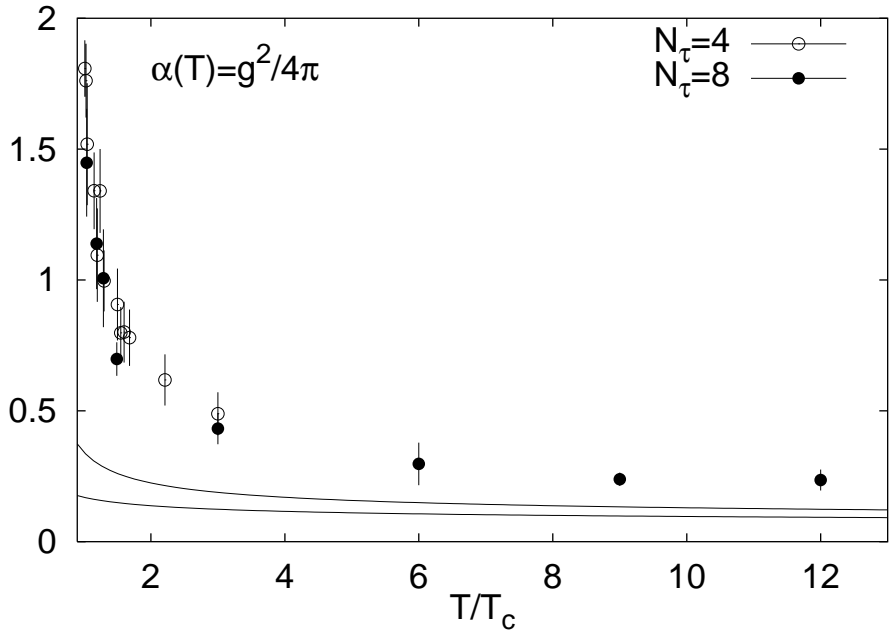,height=3.5in}}
\caption{The large distance behavior of $\alpha_S (T)$ from
evolution of the Polyakov loop in quenched LGS\cite{Zantow}.}
\label{fig2}
\end{figure}

The cutoff $\Lambda_{\rm NJL}\sim 700$ MeV may be considered as a
sort of order-parameter for the constituent quarks, as well as
Wilsonian matching scale, in that the negative energy quarks
comprise the condensate which breaks chiral symmetry in giving
hadron masses. The chiral symmetry breaking scale for hadrons is
$\Lambda_{\chi \rm SB}=\sqrt 2 \Lambda_{\rm NJL} \sim 1 $ GeV
\cite{HY:PR}. As $T$ moves above $T_c$ the (collective) chiral
symmetry breaking state disappears. The chirally restored
zero-mass $\pi, \sigma$ and $\rho$ become the thermodynamic
variables just above $T_c$. This means that the scale drops from
$\Lambda_{\chi SB}\sim 4\pi f_\pi \sim 1$ GeV towards the
infrared, the scale just above $T_c$ being zero. The large
distance color Coulomb coupling constant just above $T_c$ has been
evaluated by F. Zantow \cite{Zantow} from the Polyakov loops in
quenched LGS. We show the behavior of $\alpha(T)$ in
Fig.~\ref{fig2} for $T\gsim T_c$.

The Bielefeld $\alpha_s(T)$ does not have the Casimir $4/3$ in it
that we use. Furthermore the $\alpha_s$ is calculated with heavy
quarks. In the case of light quarks at $T_c$, addition of the
velocity-velocity interaction in the appropriate helicity states
doubles the effective Coulomb interaction. Thus our effective
large-distance $\alpha_s(T)$ would reach a value of $\sim 16/3$ at
$T_c$, giving a $g_{eff}\sim 8$ for the color Coulomb interaction.
This is a strong indication of the QCD scale having moved far
towards the infrared, indicating very low mass modes as
thermodynamic variables.

This large distance behavior does not give us the effective color
Coulomb interaction to be used in calculating our $\pi, \sigma, \rho$
bound states, which are small in extent. For these we need the
$\alpha_s (T_c,r)$ for small $r$. This is quite complicated
because we know that $\alpha_s(T,r)$ goes to zero as $r\rightarrow 0$.

We received from Olaf Kaczmarek \cite{kaczmarek} results in full
(unquenched) QCD. The results for the free energy, color singlet internal
energy, etc. are very nearly the same as in quenched QCD,
except that the $T_c$ (quenched) is rescaled to $T_c$
(unquenched). We shall assume this to be generally true in what
follows, although the vibrations have not yet been calculated in
full QCD. Note that this is full QCD for heavy quarks, so we must
add the additional effects for light quarks, such as the
velocity-velocity interaction and the instanton molecule
interaction, which would not enter into the heavy-quark sector.

We can, however, as in BLRS\cite{BLRS}, use the parameter
$\alpha_s$ obtained from the binding of charmonium above $T_c$,
namely $\alpha_s\simeq 0.5$. The charmonium atoms have about the
same radius $r\gsim 1/3$ fm as our $\sigma, \pi, \rho$ at $T_c$.
With $\alpha_s=0.5$, doubled to take into account the
velocity-velocity interaction together with the Nambu-Jona-Lasinio
interaction extended up into the chirally restored region as done
in BGLR\cite{BGLR}, BLRS obtained zero energy for their $\pi$ and
$\sigma$. Below we discuss how the $\rho$ mass is brought to zero
at $T_c$ in the chiral limit.

We remark that very strong coupling is clearly needed at $T=T_c+\epsilon$
because the $\sim$ 2 GeV unperturbed energy of the quark and antiquark
must be brought to zero by the binding potentials in the chiral
limit.

As discussed by BLRS\cite{BLRS},
the instanton molecule interactions are expected
to be unimportant by a temperature $T\sim 2 T_c$.
Note that in the chirally restored region, the scalar and pseudoscalar
degrees of freedom should be the same,
as well as those of the vector and axial vector. Thus, one can include
all excitations within the framework of the $\pi$ and the $\rho$, as
we shall do. Further note that in accord with Asakawa et al., BLRS
found the spin dependence to be negligible.

\section{The $\rho\leftrightarrows 2\pi$ Reaction as a Coupling
Between Coulomb Bound States.}
\label{sec4}

Ultimately we hope to be able to
construct the coupling between giant resonances
in the chirally restored region, but as a first step will include only
the color Coulomb interaction, which leaves out the many-particle
aspect of the giant resonances. As noted, in the chirally restored region
we need consider only $\rho$'s and $\pi$'s, which exhaust the
chirally restored degrees of freedom.

\begin{figure}
\centerline{\epsfig{file=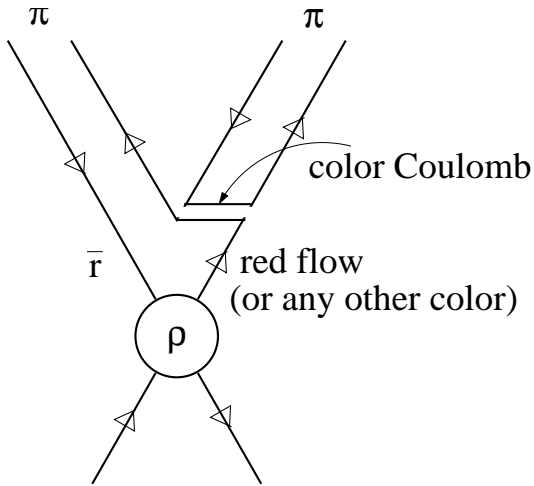,height=2.5in}} \caption{How the
color flows to show that we make colorless molecules by color
Coulomb exchange. We show only the single color Coulomb
interaction which creates an additional quark-antiquark or
quark-quark-hole pair. The quark and antiquark lines are all
Coulomb states. A factor 2 arises because the color Coulomb
interaction could equally well start from the down-going line on
the left; another factor of 2 because the two $\pi$'s could
equally contribute as two $\sigma$'s.}
\label{fig3}
\end{figure}

In Fig.~\ref{fig3} we show how a $\bar q q$ bound state with the
quantum numbers of the $\rho$ can decay into two pions.

For the transition matrix element (the matrix element to be put into
Fermi's golden rule) we have
\be
{\cal M} =\langle \bar q\gamma_5 q (1) \bar q \gamma_5 q (2)
\delta H_{23} \bar q\gamma_\mu q (3)\rangle
\ee
where the $\delta H_{23}$ is the Coulomb interaction
\be
\delta H =\frac{e^2}{r_{23}} \simeq
\frac{e^2}{r} +\frac{e^2 r_3 \cos\theta_1}{r^2}
\label{eq42}
\ee
Here $r$ is the distance between the pion at point 2, which is so small
that we take it to be a point, and the distributed mass of the $\rho$-meson
with r.m.s. radius $[\langle r_3^2\rangle]^{1/2}\simeq 0.5$ fm,
at point 3.
The distance $r$ will be $\sim 1$ fm, the average distance between particles.
We must carry the expansion to obtain the $\cos\theta$ dependence, in order
to match to the $\rho$ which we assume to be polarized in the
$z$-direction.\footnote{In fact, at $T_c$ the $\rho$ polarized in the time
direction is most important. It is related to the one in the $z$-direction
by the consistency condition $\partial\rho_\mu /\partial x_\mu =0$.
Thus, one can say that they are the same degree of freedom.}

It is of interest to compare the $\delta H$ of eq.~(\ref{eq42})
with that of the chirally broken $\rho\rightarrow 2\pi$ discussed
in Sec.~\ref{sec2}. We can obtain this from Ref.\cite{dewit}. This
is particularly simple for massless pions~\footnote{We thank M.
Prakash for this observation.} which should not be too bad an
approximation; \be {\cal M} = g_v m_\rho. \label{eq43} \ee The
ratio of the relevant part of eq.~(\ref{eq42}) to eq.~(\ref{eq43})
is \be R= \frac{e^2 r_1 \sqrt{1/3} /r^2}{g_v m_\rho}
 = \frac{4 \pi\alpha_s r_1 \sqrt{1/3} /r^2}{g_v m_\rho}
\ee
where we have replaced $\cos\theta$ by $\sqrt{1/3}$, since the square
averaged over $\Omega$ will give 1/3.

We should multiply this ratio by 4 because the Coulomb interaction could
equally well begin from the antiquark line of the $\rho$ (left-hand side
in Fig.~\ref{fig3}) and in the chirally restored regime the $\rho$ can go
into two $\sigma$'s  as well as two $\pi$'s, the $\pi$ and $\sigma$ being
the same in the chirally restored sector.

Now, as noted in the Introduction, the radius $r_3$ of the $\rho$ is
$\sim 0.5$ fm, and we found the effective $\alpha_s$ to be $\sim 1$.
We take $g_v\sim 5$. Since the particles are $\sim 1$ fm apart we take
$r=1$ fm. Note that we do not have asymptotic states in the many body
calculation, so energy conservation will be somewhat blurred.
We find $R\sim 0.18$, i.e. the nonlinearity in the $\bar q q$ Coulomb
bound states is $\sim 2/3$ as large as that of the chirally broken $\rho$
when multiplied by the 4 noted above.
However, the latter is a collective vibration, as calculated in NJL,
so before we can make a true comparison, we must carry out the same
type of loop expansion in the chirally restored region, putting in
collective effects.

At $T_c$, the effective $\alpha_s$ is about unity. As will be described
in the next section, adding loops to the $\bar q q$ bound states and
summing them will increase the process, Fig.~\ref{fig6} by a
factor $\sim 4^3=64$ at $T=T_c+\epsilon$; i.e.,
the interaction of the vibrations at $T\gsim T_c$ is much greater than
the interaction between $\bar q q$ Coulomb bound states.
Of course, what we have done is to add up an infinite number of
attractive contributions in the bubble sum. This is correct in random
phase approximation when governed by a symmetry principle; i.e.,
see \S\ 5 Ch.~V of Brown\cite{brown67} where the spurious translational
state in nuclei is brought to zero energy by just such a procedure.
The same occurs in the Anderson mode in superconductors or the
Goldstone mode $-$ our $\pi$-meson $-$ in particle physics,
except that these are real, not spurious modes.

\begin{figure}
\centerline{\epsfig{file=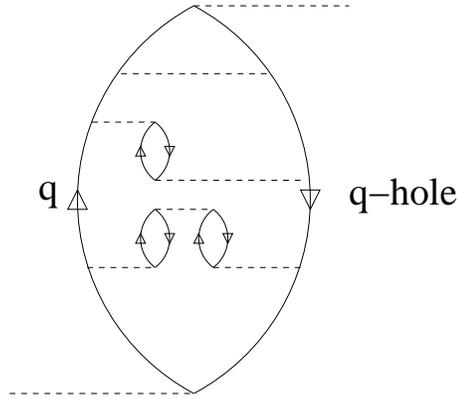,height=2.2in}}
\caption{
Higher-order graph.
In this figure we have not used Coulomb bound states as representation
but bare quark and quark-hole states, in order to make clear the
connection with the nuclear many-body problem. The dashed lines
are the Coulomb interaction. We can also attach the bubbles to the
quark or quark-hole or to each other by instanton molecule interactions.
}
\label{fig4}
\end{figure}

However, in the part of the nuclear many-body problem not governed
by symmetry principles, the original Kuo-Brown procedure
of including only one bubble has turned out to be quite good
\cite{Kirson71}. Of course our situation here is different from the
nuclear many-body problem, in that we have both the Coulomb interaction
and bubbles. However, the general idea may be the same.
In Fig.~\ref{fig4} we show a typical higher order graph summed
by Kirson in his ``bubbles in bubbles" translated to our
Coulomb$+$instanton molecule interactions.

The ``bubbles in bubbles" increase the contributions of repulsive
terms -- the process shown in Fig.~\ref{fig4} is an exchange
term -- in higher order. Once the interactions are more accurately
determined; e.g. by LGS, it will be interesting to attempt to sum
them. (In the case of the nuclear many-body problem, even though it is
a case of strong interaction, such a systematic treatment has been
carried out\cite{Kirson71}.) However, for the moment we will introduce
just one bubble, assuming the higher-order bubble terms to be cut
down by the exchange terms with bubbles in bubbles.

\begin{figure}
\centerline{\epsfig{file=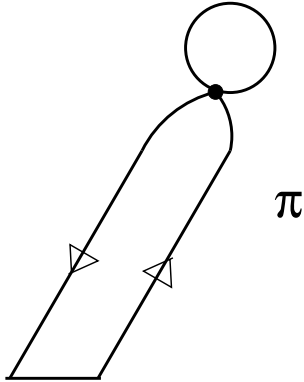,height=2.0in}}
\caption{
Adding a bubble to one of the $\pi$'s in Fig.~\ref{fig3},
which also involves adding a propagator. The solid dot is the
instanton induced interactions. All lines are Coulomb bound states.
}
\label{fig5}
\end{figure}

{}From the development Fig.~5 and following in BLRS\cite{BLRS} we note
that adding a bubble as in Fig.~\ref{fig5} will increase the
contribution by the factor
\be
C=1+\frac{1\; {\rm GeV} \; F}{1.5\; {\rm GeV}} =\frac 43
\ee
where $F\simeq 0.5$ is the correction for nonlocality.
(See Eq.~(42) and Appendix of BLRS\cite{BLRS}).
Now the bubble can be added to either $\pi$ or the
initial $\rho$ in Eq.~\ref{eq43} so that the overall increase is by a
factor of $(1+1/3)^3 \simeq 2.4$.

Including the factor 2.4 in $R$ we find the Golden rule matrix element
$M$ to be 1.6 times that for the chirally broken
$\rho\leftrightarrows  2\pi$, or the width to be a factor of
$(1.6)^2$ larger, i.e.
\be
\Gamma=380\; {\rm MeV}.
\label{eq6}
\ee
Whereas the above is hardly a quantitative calculation, we do end
up with a width $\Gamma$ somewhere in between the $m_\rho$ of 560 MeV
of BLRS and the 280 MeV we shall end up with here. This is satisfying in that
in strong coupling the width of a particle should end up more or
less equal to its mass.

In order to come down to the one-bubble insertion as taking care of
the role of vibrations in the nuclear many-body problem, a lot of
calculations, ending in Kirson's work\cite{Kirson71} had to be
carried out. In our present situation we have both color Coulomb,
which scales rapidly with energy, and the instanton molecule
interaction which is tied to the hard glue. Thus a quantitative
many-body calculation would be difficult. We do, however, have the
lattice calculations. Dealing with heavy quarks these do not
include the velocity-velocity and instanton molecule interactions, but
this may be an advantage for investigating how the widths of the
resonances grow as $T$ moves downwards towards $T_c$, in that they
should not get out of hand, which they do with our estimates
including everything.

Given the fact that the coupling must be so strong as to bring the
unperturbed (chirally invariant) mass $2 m_q$ down to zero in
order to construct the $\pi$ and $\sigma$ in the chiral limit, we
believe that they will give a strong $\rho\leftrightarrows 2\pi$
transition. Since we have only $\rho$'s and $\pi$'s in the
chirally restored region above $T_c$, this will result in complete
equilibration. (It should be noted that below -- but away from --
$T_c$ the $\rho\leftrightarrows 2\pi$ reaction is the strongest,
and results in equilibration of the $\rho$ with the two pions down
to thermal freezeout.)

We finish this section by showing that the same sort of strong
interactions that produced the $\rho\leftrightarrows 2\pi$ nonlinearities
will give large meson scattering cross sections. We draw the color
flow in Fig.~\ref{fig6} for pion-pion scattering.
In the case of the pions, we would use the first term $e^2/r$
of $\delta H$, eq.~(\ref{eq42}) and it is easy to see that we will
get cross section of hadronic size, since $\alpha_s\sim1$ at $T_c$.

\begin{figure}
\centerline{\epsfig{file=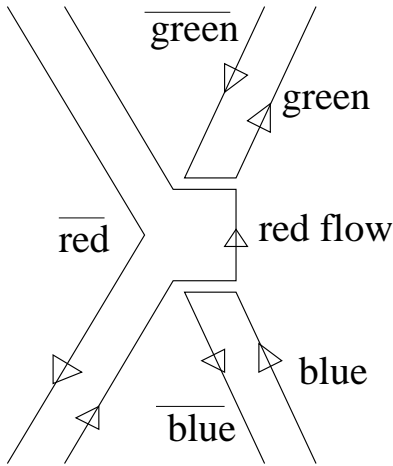,height=2.5in}}
\caption{
Pion-pion scattering in the Furry representation. The red-$\overline{\rm blue}$
and green-$\overline{\rm red}$ contributions to the color Coulomb interaction
are shown; of course, the color must be summed over. The other
color Coulomb (singlet) interactions which establish the Furry
representation are not shown.
}
\label{fig6}
\end{figure}

\section{Constructing Vibrations by connecting Coulomb Bound States}

Now the instanton molecule interaction for $T\gsim T_c$ is
sizable, and effects in producing giant collective modes are
maximized by the degeneracy in unperturbed energies at the common
$1.5$ GeV energy of the Coulomb bound states. Because of the
degeneracy in unperturbed state energy, the problem for $T>T_c$
actually resembles more that of the giant dipole state in nuclei
\cite{brown49} than NJL for $T<T_c$. From Ref.\cite{brown67} one
can see that if one includes only loops going forward in time,
then the energy of the giant collective mode is brought down with
the entire trace of the secular matrix for these states. The
argument is repeated in Appendix A. Furthermore, if the forward
going loops bring the energy of the collective mode halfway from
the $1.5$ GeV unperturbed energy to 0.75 GeV, then including the
backward going loops (``ground-state correlations") will bring the
energy all of their way to zero\cite{brown67}. In other words, all
of the interactions conspire so as to move the collective mode in
energy, leaving the remaining non-collective modes at their
unperturbed energy of 1.5 GeV.
(As noted earlier, the fact that the pion is a Goldstone mode
governs the attraction from all sets of bubbles.)

Indeed, with such strong interactions we will certainly have
thermal equilibrium in the region just above $T_c$
(and probably higher).

Although not yet confirmed by LGS,\footnote{ The necessary
unquenched calculations have not been carried out.} the scenario
by Harada \& Yamawaki \cite{HY:PR} in which the width of the
$\rho$-meson goes to zero (in the chiral limit) as $T\rightarrow
T_c$ from below, should be helpful in providing equilibrated
hadron emission. In their scheme $g_V\rightarrow 0$ as
$T\rightarrow T_c$ from below. In fact, Brown and Rho
\cite{brown02} showed that ${g_V^\star}^2/{m_\rho^\star}^2$ goes
smoothly through $T_c$ from LGS of quark number susceptibility.
Thus if $m_\rho^\star\rightarrow 0$, $g_V^\star$ does also.
Furthermore, the $G$ in BGLR\cite{BGLR} is essentially the ratio
$g_{\sigma QQ}^2/m_\sigma^2$ and the rough constancy in $G$ means
that as $m_\sigma$ goes to zero at $T_c$ in the chiral limit, so
does $g_{\sigma QQ}$.

Thus, the system of highly equilibrated chirally restored mesons
just above $T_c$ cut loose at $T_c$ into an environment in which
the interactions are zero in the chiral limit
just below $T_c$. This is
obviously the optimum condition for chemically equilibrated
particle yields at freezeout, which takes place at $T_c$, as
foreseen by Braun-Munzinger et al.\cite{BMSW03}.


\section{Construction of $F=E-\tau\sigma$.}

The Helmholtz free energy $F=E-\tau\sigma$ enters into the
exponent of the Wilson line as $\exp(\beta F)$ and is
proportional to the string tension so $F=0$ at $T_c$ for
heavy mesons. In BLRS we found it extremely useful to use
the Bielefeld lattice results for heavy mesons in charmonium,
and to add the additional effects relevant for light
quarks such as those from the Ampere's law velocity-velocity
interaction and the instanton molecule interaction.
Recently there has been a lot of work at Bielefeld
evaluating the Helmholtz free energy for heavy quarks \cite{Bielefeld}.
It is
of interest to calculate $F$ for our chirally restored mesons,
which add some new aspects,  here.

In the appendix we give the solutions for the instanton molecule
Lagrangian of BLRS. For any given spin and isospin that the
Coulomb bound states are coupled to, they form an unperturbed
representation degenerate in energy, each $\bar q q$ bound state
lying at 1.5 GeV. All matrix elements, both diagonal and
off-diagonal, between unperturbed states are equal, given by the
4-point (zero range) instanton molecule interaction constructed by
BLRS. For each channel, i.e., given spin and isospin, one
collective state moves down to zero energy and all the other
eigenstates remain at their unperturbed position of 1.5 GeV. Now
the partition function is the sum of Boltzmann factors
$Z(\tau)=\sum_n \exp (-\epsilon/\tau)$. For the moment we consider
only momentum zero states. The zero-energy collective state will
contribute unity to the sum. The other states will each contribute
$\sim \exp (-1.5 {\rm GeV}/0.175 {\rm GeV})$ at $T=T_c=175$ MeV,
or $\sim 2\times 10^{-4}$.

We thus approximate $Z$ for a boson at rest, by
\be
Z=\sum_n e^{-\epsilon_n/\tau}=1,
\ee
keeping only the zero energy collective state. Thus, the free energy
is
\be
F=- T \ln Z =0.
\ee

All that this shows is that in the construction of our chirally
restored mesons, the large necessary binding energy reduces the
large $\gsim 1$ GeV quark masses to zero, affecting neither
pressure nor entropy, given by \be p=-\frac{\partial F}{\partial
V},\;\; \sigma= - \frac{\partial F}{\partial\tau}. \ee

At temperatures well above $T_c$, where the vibrations are linear,
and the epoxy is melted -- roughly $T=2 T_c$ as will be
discussed in the Appendix, we have, to a good approximation,
a Boltzmann gas, with each boson carrying $\epsilon\simeq 2.7 T$
and $\sigma=3.6$. The free energy is obviously a minimum.

However, the connection between equilibration and minimization of
the free energy is not clear at $T_c$, and possibly not at $1.4
T_c$, with the vicinity of which we are concerned. The usual
derivation involves differentiating $F$ to find the extremum \be
dF= dE -\tau d\sigma \ee and showing that the right hand side is
zero in thermal equilibrium. However, at $T_c$ the bag constant
$B$, which describes the energy in the epoxy, about half of the
total bag constant (the soft glue being melted by $T_c$) increases
the energy density without changing the entropy. It thus
contributes negatively to the pressure. For many years lattice
gauge simulations tended to find the pressure to be negative at
$T_c$, which was, of course, unacceptable because the system would
collapse. Brown et al.\cite{BJBP93} in a crude calculation got the
chiral restoration temperature to be $T_c=172$ MeV just by
requiring the restored phase to begin as soon as the pressure
could be made positive.

Although we don't know whether the Boltzmann gas applies to the region
around $T_c$ where the vibrations are very nonlinear, we use it in
Sec.~\ref{discussion} to discuss the density of the thermal excitations
which must be added to bring the pressure to zero.

In the Harada and Yamawaki scheme\cite{HY:PR} it is easy to see why
the pressure in the hadron gas is low at $T_c$, because the interactions
go to zero in the chiral limit. (Of course pressures must be equal in
both phases in a phase transition.) In the resonance gas\cite{Redlich}
the pressure on the hadron side is made small by putting most of the
energy into the binding energy of excited states.

We can only conclude that the free energy probably is close to zero
in the chirally restored phase at $T_c$, the value that results
for heavy quarks from the Wilson loop.


\section{The $\rho$-meson is special}

As noted above, because of chiral restoration, we have only two
types of chirally restored mesons $\pi$ and $\rho$ above $T_c$.
The chirally restored $A_1$ is equivalent to the $\rho$ and the
chirally restored $\sigma$ is equivalent to the $\pi$.
Furthermore, because the $\vec\tau$ in the instanton molecule
interaction is a four-vector, and all but the last term in the
instanton molecule Lagrangian in BLRS\cite{BLRS}, to which we
return in discussion below, involve the square of $\vec\tau$, the
different isospin states are degenerate. In BLRS\cite{BLRS} we
showed this to be nearly true for spin effects, which were
negligible. Thus, we are left with only the $\rho$ and $\pi$, with
their various spins and isospins, in an SU(4) multiplet.

In the chiral limit the $\pi$ is constrained as Goldstone boson to
have zero mass at $T_c$ (just below and just above). Therefore,
only the mass of the $\rho$ can be abjudicated; i.e., the only
model dependence on meson masses at $T_c$ comes in the
$\rho$-mass. By $T_c$ here we mean coming down to $T_c$ from
above; we know that going up to $T_c$ below there is a fixed point
in the $\rho$-mass of zero at $T_c$ \cite{HY:PR}.

Although the mass of the $\rho$ was found in classical approximation
to be zero in BLRS\cite{BLRS}, the instanton molecule fluctuated
quantum mechanically about the time axis, with r.m.s. fluctuation
in $\theta_4$, the angle with this axis, of about $30^\circ$, so that
$m_\rho$ ended up at 560 MeV. However, our strong $\rho\leftrightarrows 2\pi$
coupling of Sec.~\ref{sec4} means that the lowest-lying eigenmode in the
$\rho$-channel will be a ``rhosobar"; in this case, a coherent linear
combination of $\rho$ and $2\pi$, essentially
\be
|{\rm rhosobar}\rangle = a|\rho\rangle + \sqrt{1-a^2} |2\pi\rangle.
\ee
Consider the second-order self energy of the $\rho$ from this coupling
shown in Fig.~\ref{fig7} below.
Of course this mixes the various $2\pi$ states only perturbatively with
the $\rho$ which is inadequate for a quantitative estimate.

\begin{figure}
\centerline{\epsfig{file=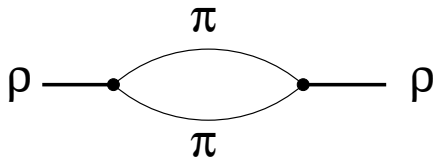,height=1in}}
\caption{Second order self energy contribution to the $\rho$ mediated
by $2\pi$ exchange.}
\label{fig7}
\end{figure}

If $m_\rho$ is greater than $2 m_\pi$, then this loop will involve a
principal value integral with both positive and negative contributions.
Thus, the greatest attraction is obtained when $m_\rho=2 m_\pi$.
Thus, in the chiral limit $m_\rho\sim 0$, but with chiral symmetry breaking,
$m_\rho\simeq 280$ MeV, twice the pion mass.
We do not consider this a rigorous argument, but show below that
the result is reasonable in the sense that going just below $T_c$
where it is emitted the $\rho$ gets its on-shell mass back.
This procedure also minimizes the free energy (under the condition
that the $\rho$ has the chirally broken mass of $2 m_\pi$)
which should be achieved for equilibrium.
We illustrate by this argument that chiral symmetry breaking will be
very important in determining the mass of the $\rho$.
Of course this sharp $m_\rho$ is used only in lowest-order
estimates. With the 380 MeV width we had in eq.~(\ref{eq6}) the
$\rho$ strength will be spread widely up to nearly 500 MeV.

Note that there is an equal degeneracy of longitudinal $\rho$'s as pions
at $T_c$. Therefore, if the $\rho$ has a mass of $\sim 280$ MeV and
we neglect the $\sim 140$ MeV difference of this mass from the pion
mass, we would ultimately, as the temperature decreased towards thermal
freezeout, get about as many pions from the decay of the longitudinal
$\rho$'s as from the pions themselves. This would be the situation
if all interactions cut off below $T_c$, as envisioned by Harada
and Yamawaki\cite{HY:PR} where the width of the $\rho$ goes to zero
in the chiral limit. But, of course, this is not so good an approximation,
especially not at $T_c$, where our estimates for $T_c+\epsilon$
are mixed with those of BSW\cite{BMSW03}
for $T_c-\epsilon$ through the explicit
breaking of chiral invariance, which converts a phase transition into
a smooth crossover transition.

In fact, Braun-Munzinger et al.\cite{BMRS03} note a substantial
discrepancy in the $\rho_0/\pi^-$ ratio in semi-central Au-Au
collisions. We give their discussion in what follows: These mesons
have been reconstructed in STAR \cite{STAR} via their decay
channel into 2 charged pions. Comparing the preliminary results
from STAR with the thermal model predictions of BRS\cite{BMRS03},
reveals that the measured values exceed the calculated values by
about a factor of 2. This is unexpected, especially considering
that BRS use a chemical freezeout temperature of 177 MeV for the
calculation. One might expect these wide resonances to be formed
near to thermal freezeout, i.e. at a temperature of about 120 MeV.
At this temperature the equilibrium value for the $\rho_0/\pi^-$
ratio is much smaller than the 0.11 found at 177 MeV. Even with a
chemical potential for pions of close to the pion mass and taking
into account the apparent (downwards) mass shift for the $\rho_0$
it seems difficult to explain the experimentally observed value of
about 0.2. In fact, recently the STAR paper was published
\cite{STAR2} where they give the final $\rho_0/\pi^-$ ratio as
$0.169\pm 0.003 (stat) \pm 0.037 (syst)$ for peripheral Au$+$Au
collisions, slightly lower than the preliminary $\sim 0.2$. Other
statistical models \cite{rapp03,Broniowski03} also have difficulty
in explaining this large a ratio. We return to this below.

Now this excess is difficult to explain with the $\rho$ width
that of the free $\rho$, $\Gamma_\rho\sim 150$ MeV, during the
drop in temperature from 175 to 120 MeV. This gives the strongest
hadronic interaction and would surely be large enough to equilibrate
the $\rho$'s and $\pi$'s down to thermal freezeout, the
$\rho$'s changing back and forth between two $\pi$'s several times.
Since they are
obviously not equilibrated, the assumption that the $\rho$-meson
has its free-particle width must be wrong, and we adduce the
Harada and Yamawaki argument to say that only that part of the
$\Gamma$ which originates from explicit chiral symmetry breaking
remains at $T_c$; in any case, that the in-medium width is
substantially smaller than the free-particle one.

Let us try the extreme version of the theory in which the
$\rho$ and $\pi$ mesons stop interacting as they drop below
$T_c$ in temperature.
In Appendix C we develop  a schematic model, which has
the important ingredients, for how the $\rho$ of mass $2 m_\pi$
at $T_c$ goes on shell before leaving the system.
Out of the $\rho_0$'s, only the longitudinal
one will be of low energy, because of the instanton-molecule
polarization, as explained in BLRS\cite{BLRS}. Its strong
attractive principal value interaction with the pion is maximized
in magnitude by a mass of $m_\rho=2 m_\pi$
(so that there are no negative contributions to the principal
value integral). We believe the nonzero longitudinal $\rho$ mass
to be the main effect of explicit chiral symmetry breaking.
We use Boltzmann factors in our schematic estimates.

%
%
%
The total energy, rest mass plus thermal, of the 770 MeV $\rho$
at $T_c=175$ MeV is 1090 MeV, so the Boltzmann factor multiplied by
3 for the spins is 0.006, in the standard scenario. In our case only
the longitudinal $\rho$ has the low mass of 280 MeV at $T_c$.
Its total energy is 652 MeV, with Boltzmann factor of 0.024.
Thus, even though the degeneracy is cut down by 1/3, because of
its lower mass, its abundance would be 4 times greater. Thereafter,
in the chiral limit with $\Gamma_\rho=0$, the $\rho$'s would just
leave. Given that the explicit chiral symmetry breaking is present
and that the $T_c+\epsilon$ region will be overlapped with the
$T_c-\epsilon$ region in a smooth crossover transition (see the
next section) we would expect the abundance to be substantially
decreased, but still present.
The experimental enhancement by a factor $\sim 2$, half of our
enhancement in the chiral limit, looks reasonable.

As the longitudinal $\rho$ goes below $T_c$ in temperature, it will
gain back $\sim$ 0.33 GeV binding energy it got from the color
Coulomb interaction just above $T_c$, going on shell.\footnote{
We take the $\alpha_s$ below $T_c$ to be that of charmonium,
$\alpha_s\sim 0.33$. Just above $T_c$ where the Coulomb binding is
$\sim 0.5$ GeV (BLRS\cite{BLRS}, Table 1), the effective
$\alpha_s$ is $(\alpha_s)_{\rm eff}\simeq 1$.
So, the 0.33 GeV is 2/3 of the 0.5 GeV.
}
The reconstituted spin interaction also adds to the mass,
$\sim 0.25 (m_\rho -m_\pi)=0.16$ GeV. The addition of these
to the $\rho$-mass of $0.28$ GeV at $T_c+\epsilon$ puts
the $\rho$ on shell at $T_c-\epsilon$.

Although our calculation is made only within SU(2), we believe
that the spin effects will also be substantially weaker above
$T_c$ than below in SU(3). Now the ${\bar K}_0^\star$-$K$
splitting below $T_c$ is only about half of that of the
$\rho$-$\pi$ splitting, so the enhancement should not be so strong
in the ${\bar K}_0^\star$ abundance as in that of the $\rho$, but
a factor of $\sim 2$ in the ${\bar K}_0^\star$ abundance over that
determined by equilibration at $T_c-\epsilon$ would only improve
matters \cite{BMRS01}, although the difference between prediction
and experiment is only one standard deviation in
Ref.\cite{BMRS01}.

We did not find large deviations from the standard scenario in the
STAR $K^{\star 0}/K$ and $\phi/ K^{\star 0}$ ratios. We can see
that medium effects will decrease both numerator and denominator
here.

In our work we have relied upon chiral invariance to construct the
$\pi$ and $\sigma$ so that their masses go smoothly through chiral
restoration. We have not considered baryons and don't know how to
introduce them. We do know that the $\Delta$ cannot have got much
of its 300 MeV mass relative to the nucleon from the perturbative
spin-spin interaction as in the MIT bag model. This would be
absent about $T_c$, with the result that numbers of nucleons
and $\Delta$'s would be equal there. Whereas STAR shows some
excess of $\Delta$'s above the standard equilibrium scenario
at central rapidity, this excess of $\sim 1/3$ is small compared
with the $\exp (300 {\rm MeV}/T_c)$ that would be obtained
in the perturbative scenario. More likely is that the singlet,
isosinglet diquark which is strongly bound, and which exists
in the nucleon, but not in the $\Delta$, gives most of the
energy difference between (possibly incipient) $\Delta$'s and
nucleons above $T_c$. The binding of the diquark, as our Nambu-Jona
Lasinio interaction, goes smoothly up through $T_c$ since it
results from the 't Hooft instanton interaction. Above $T_c$
it will also have an attractive color singlet interaction.
Thus, in some way the nucleon may be formed above $T_c$ by the
diquark picking up a quark, but the $\Delta$ would have to be
formed by the coalescence of three quarks. In any case, the difference
in energy between the incipient $\Delta$'s and nucleons just above
$T_c$ cannot be much smaller than just below $T_C$.

Of course, ratios of antiparticle to particle masses will be
unchanged by whether equilibration is just below or just above $T_c$.

There are, thus, lots of ways in which the role of the medium
dependence in equilibrium is hidden. It is no surprise that its
role is clearest in the $\rho/\pi$ ratio, because most of the
$\rho$ mass is generated dynamically in the chirally broken
sector, whereas the pion, as a Goldstone boson, remains unaffected
(except for the explicit chiral symmetry breaking). It is just
this behavior of $\pi$ and $\rho$ which led to Brown-Rho
scaling\cite{br91}, which was first discovered empirically in
nuclear spectra, because the $\pi$ and $\rho$ contribute to the
tensor force with opposite signs. Thus, as the $\rho$-mass
decreased with density, the $\rho$ contributed more strongly and
weakened the tensor force \cite{BR03}.



We remark briefly that our above discussion does not include the
last term \be \delta L_{\rm IML} = - (\bar\psi \gamma_\mu\gamma_5
\psi)^2 \ee of the instanton molecule Lagrangian of eq.~(37) of
BLRS\cite{BLRS}. Above $T_c$ with chiral restoration this is the
same as \be \delta L_{\rm IML} = (\bar\psi\gamma_\mu\psi)^2; \ee
i.e., attractive coupling of the $\omega$-meson. This may be
connected with a discontinuity in the baryon number chemical
potential~\cite{isb}.

\section{The Braun-Munzinger, Stachel and Wetterich (BSW) Scenario}

Our scenario is that equilibration takes place just above $T_c$ in
the chirally broken sector. We have constructed a strongly
interacting colorless liquid in that sector. Nonetheless, our
scenario has important points in common with that of
BSW~\cite{BMSW03}, although they consider equilibration to take
place in the chirally broken hadronic sector just below $T_c$.

These authors make 3 main points:
\begin{enumerate}
\item Dominance of hadronic
reactions with a high number of ingoing particles can be
realized only very close to the phase transition.
\item Two-particle processes are too slow to establish and sustain
chemical equilibrium near the chemical freezeout temperature
$T_{ch}$.
\item Multi-particle scattering is indeed fast enough in order
to maintain equilibrium for $T\ge T_{ch}$.
\end{enumerate}

These authors carry out an extremely illuminating calculation of
the rate $r_\Omega$ for $\Omega$ production, through the reaction
$2\pi+3 K\rightarrow \bar N\Omega$. As we, they take perturbative
(Boltzmann) thermal model densities. For the Golden rule matrix
element they take the measured $p+\bar p\rightarrow 5 \pi$ cross
section. They check by evaluating the rate for $\Omega$
production\footnote{ Because of the triple strangeness of the
$\Omega$, it may be considered the most difficult particle to
equilibrate.} in a semi-classical approach, in which the standard
two-body rate equation is generalized to multi-particle
collisions, obtaining essentially the same result.

\begin{figure}
\centerline{\epsfig{file=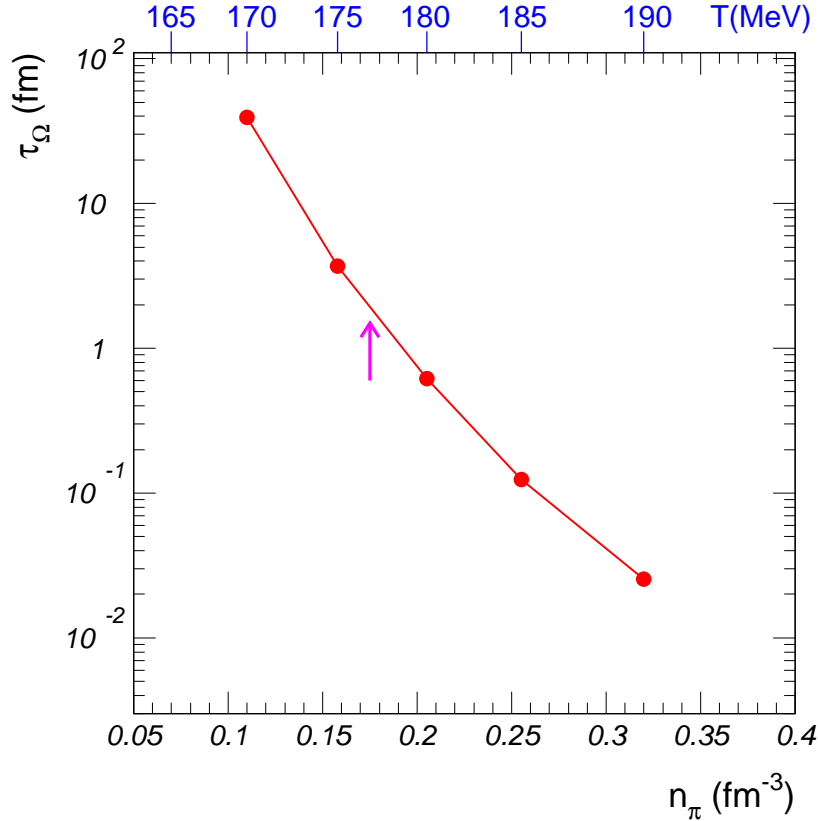,height=5in}}
\caption{Time $\tau_{\Omega} = n_\Omega/r_\Omega$ needed
to bring $\Omega$ baryons
into chemical equilibrium via multi-particle collisions.
The arrow points to the Boltzmann pion density at $T_c$.
}\label{fig8}
\end{figure}

Thus BSW reach a time of 2.3 fm, sufficiently short to bring the
$\Omega$'s into equilibrium. Close to $T_c$ the $\Omega$ equilibration
scales approximately as $r_\Omega\propto T^{-60}$~!

Therefore, in a standard hadronic environment with reasonable parameters,
chemical equilibration can be determined close to $T_c$, say
at $T_c-\epsilon$.

In our paper above we have worked at $T_c+\epsilon$, i.e., in the
chirally restored phase, from a different scenario, but with the
same general results. Although as we formulated it, in the chiral
limit, our scenario looks very different from that of BSW, we note
that introducing the explicit chiral symmetry breaking would
smooth our sharp chiral restoration transition into a smooth
crossover one, mixing what happens in our scenario at
$T_c+\epsilon$ to $T_c-\epsilon$. Also, even granted the
Harada-Yamawaki results in the chiral limit\cite{HY:PR}, their
$\rho$-meson width will not be zero with the explicit chiral
symmetry breaking. Furthermore, with their $T^{-60}$ factor for
the $\Omega^-$, equilibration would be likely to continue to
$T_c-\epsilon$.

In practical terms, i.e., fitting the data, the BSW work has been
extremely successful. Nonetheless, there may be exceptions such as
we discuss with the $\rho$-meson, which cannot be explained in
their work.

We note that inclusion of the $T_c+\epsilon$ region in their
calculations can only help them. At $T_c+\epsilon$ we have 8
pions, as compared with 3 below $T_c$, the $\sigma, \rho$,
isosinglet pseudoscalar and pions all being equivalent in the
chirally restored region. The $\tau_\Omega$ in Fig.~\ref{fig8}
goes with the inverse 5th power of the pion density, so the higher
degeneracy just above $T_c$ would be of help.

Let us turn the previous statement around and say that the BSW
work can help us. Namely, with the greater number of pions and the
stronger interactions the material just above $T_c$ is clearly
well and truly equilibrated.

We believe that our scenario has the following merits:
\begin{enumerate}
\item Chirally restored particles cannot leave the fireball. Therefore
the latter must cool until they can come down to $T_c$.
\item However, equilibration at $T=T_c+\epsilon$ and the
BSW equilibrium at $T=T_c-\epsilon$ which increases so strongly
with $T$ that even if zero in the chiral limit, is likely to be
large because of explicit chiral symmetry breaking, will be mixed
by the smooth crossover transition, over the region
$\epsilon \sim 5 $ MeV, the region below $T_c$ in which BSW find
equilibration. Thus, this BSW ``band" will be the region in
which thermally equilibrated mesons emerge.

\end{enumerate}

We believe that the BSW paper and our note are the beginning of an
unravelling of the puzzle of equilibration.

\section{Discussion}
\label{discussion}

Brown et al.\cite{BJBP93} suggested that the mesonic resonance
states $\pi, \sigma, \rho, A_1$ go smoothly through $T_c$ with
increasing temperature, simply changing from being chirally broken
to chirally restored. The same general idea is retained in the
resonance gas picture \cite{Redlich} in which resonances in
addition to the above hadrons are included so as to make the
transition smooth. As noted in Sec.~\ref{sec2}, the $\pi, \sigma,
\rho, A_1$ can be constructed as vibrations in the
Nambu-Jona-Lasinio language below $T_c$ so that our present
scenario is not new in making them vibrations above $T_c$.
However, it is easier to do the thermodynamics above $T_c$,
because the $\bar q q$ molecules form a degenerate unperturbed
configuration, so that all of the strength ends up in the
collective state, none in the non-collective ones. This degeneracy
endows the $\gsim T_c$ region with greatly enhanced interaction
strengths.

The pressure must be nonnegative at $T_c$. This is achieved in the
resonance gas picture by introducing mesonic excited states as well as
resonances, as well as other hadrons than we consider.

Indeed, at $T_c$ we will need additional degrees of freedom because
in order to bring the $\rho$ and $A_1$ masses to zero we must polarize
them in the time direction, and this leaves us with only 8 degrees
of freedom, which together with the $\pi$ and $\sigma$ give 16 at $T_c$.
As noted, with this degeneracy we have only $\sim 1$ boson/fm$^3$
at $T_c$, which in Boltzmann approximation gives a pressure of
$p=0.9 T_c/{\rm fm}^3 = 158 {\rm MeV/fm^3}$, a bit higher when
calculated in detail for bosons. Now in BGLR\cite{BGLR} we found
that $\sim$ half of the $\sim 500$ MeV bag constant was melted
in the soft glue by $T=T_c$, $\gsim$ half remaining in the epoxy,
or hard glue. The $-250$ MeV/fm$^3$ contribution from this to the
pressure leaves us with a deficit of $\sim 100$ MeV/fm$^3$.

A part of the pressure lost from the transverse $\rho$'s and $A_1$'s
being high in mass will be filled in by pressure from the glueballs.
The soft glue that is melted as $T$ goes up to $T_c$ will go into
glueballs, but these are high in mass below $T_c$, although certainly
not higher than the $\sim 1.5-2.0$ GeV masses of glueballs at zero
density.
The Casimir operator is 3 for gluon pairs, compared with $4/3$ for
quarks, and so the color Coulomb should bring the $gg$ bound states
down by 1.25 GeV. Petreczky et al. find thermal gluon masses to be
$\sim 20\%$ lower than thermal quark masses at $T=\frac 32 T_c$
and $T=3 T_c$. With our assumed 1 GeV for the
thermal quark mass, this would bring the $g g$ bound states
down to the region of masses where, with a multiplicity of 8 they could
bring the pressure most of the way to zero.

Some pressure will come from the kaons: $K^+, K^-, K^0, \bar K^0$
and the vector\footnote{ In fact the vector masses may be
considerably less, resembling to some extent the $\rho$-meson.}
$K^\star (892)$ whose masses at the SU(2)$\times$SU(2) $T_c\sim
175$ MeV should be somewhat decreased below the free masses by the
melting of the non-strange quark condensates and partial melting
of the strange quark condensates.

Hadrons other than our collective states will enter in and there will
be some contribution from excited states of our chirally restored
$\pi, \rho, \sigma, A_1$ although these states will lie at high
energies because of the smallness of our molecules.

We see no problem in obtaining sufficient pressure well above $T_c$,
say at $2 T_c$, because the remainder of the bag constant will go as
the epoxy is melted and our degrees of freedom will double to 32
as the $\rho$ and $A_1$ are no longer polarized.

\section{Conclusion}

We have argued for equilibration of our excitations just above
$T_c$. Every quark and antiquark within the relevant rapidity
interval participate equally in a set of SU(4) vibrations, the
energies of which go to zero in the chiral limit at $T_c$. At a
higher temperature, approximately that reached in RHIC following
the color glass stage ($t \sim \frac 23$ fm/c), those vibrations
are found in the lattice gauge calculations of Asakawa et
al.\cite{asakawa03} and Wetzorke et al.\cite{wetzorke}.

The mesons remain equilibrated, at least in the chiral limit, as
they are emitted, since the interactions have been found by Harada
and Yamawaki\cite{HY:PR} to go to zero as $T$ goes up to $T_c$
from below. And they cannot be emitted until $T$ has come down to
$T_c$ from above. In the more realistic chirally broken system, we
believe that chemically equilibrated mesons will emerge in the
$\sim$ 5 MeV band below $T_c$ of Braun-Munzinger, Stachel and
Wetterich \cite{BMSW03}.

Whereas the zero masses of $\pi$ and $\sigma$ are protected by
chiral symmetry at $T_c$, both slightly above and slightly below,
the $\rho$-meson mass must go on shell at 770 MeV as it materializes,
even though the abundance of $\rho$'s is determined when its mass
is nearly zero, slightly above $T_c$. In the chiral limit this means
that the number of $\rho$'s, which are reconstructed from their
$2\pi$ decay, just below $T_c$ should be
greater than predicted by Boltzmann factor using their 770 MeV
on-shell mass.
A substantial excess in this number, which appears anomalous in
the conventional treatment, should remain even after introduction of
bare quark masses.

Interactions which dictate the above picture are strong with $g\gg 1$,
very different from those in the perturbative quark-gluon plasma
often predicted for RHIC energies.

We believe that we have established, backed by the LGS of Asakawa
et al.\cite{asakawa03}, that the matter formed by RHIC is composed
of giant collective vibrations of $\bar q q$ pairs and powered by
the instanton molecule interaction. In a particle-wave duality
that existed in the work of Nambu-Jona-Lasinio, these vibrations
can also be interpreted as mesons. Above $T_c$, these mesons are
chirally restored. In the formation of these colorless mesons,
color is dynamically confined somewhat above $T_c$.

Normally one associates the large and rapid increase with
temperature of the entropy $s$ in the region of $T_c$, with the
formation of the quark gluon plasma. However, Koch and Brown
\cite{Koch1993} (see especially Fig.~3) showed that hadrons going
massless provided an increase in entropy that matched that found
in LGS. In fact, somewhat above $T_c$ (unquenched) where the
$\rho$ and $A_1$ are no longer polarized in the time direction, we
have 32 degrees of freedom in the instanton molecule scenario. To
the extent that these are essentially massless, these 32 bosonic
degrees of freedom provide the pressure found in the SU(2) lattice
gauge calculations \cite{Karsch02}. Thus, one might say that the
large increase in number of degrees of freedom results from the
hadron masses going to zero in the chiral limit; i.e., Brown-Rho
scaling~\cite{br91}.

Thus, RHIC has found a new kind of matter, one we find very
interesting. In the sense that mesons are composed of quarks and
gluons, it may be simple-mindedly called ``quark-gluon plasma,"
but hardly so in the sense of the $predicted$ perturbative QGP.

This same new form of matter must have been gone through
in the early universe, as $T$ decreased from $\sim 2 T_c$
to $T_c$. The nice, smooth, essentially
second-order transition we construct would mitigate against
inhomogeneities originating in the
chiral restoration transition.
Since we have dynamical confinement just above $T_c$, we have not had
to discuss deconfinement, which presumably requires a discussion of
Polyakov lines.

\section*{Acknowledgments}
We are grateful to Peter Braun-Munzinger and Johanna Stachel for
many discussions over the years in which they pursued the
equilibration at chemical freezeout. We are grateful to Felix
Zantow for the lattice gauge results from his thesis and to Olaf
Kaczmarek and Peter Peter Petreczky. G.E.B. has had many fruitful
discussions with Edward Shuryak to whom he is particularly
grateful for having stressed for more than a decade movement
towards the ``softest point" below $T_c$, for discussion of the
scenario presented in Appendix B and for undertaking the niche
into which the Harada-Yamawaki HLS/VM theory fits. GEB was
supported in part by the US Department of Energy under Grant No.
DE-FG02-88ER40388. CHL was supported by Korea Research Foundation
Grant (KRF-2002-070-C00027).


\appendix
\section{Appendix A: Comparison with Lattice Gauge Simulation}
\label{appA}

As noted in BLRS\cite{BLRS} the main idea of Shuryak and Zahed
\cite{shuryak2003} was that ``after deconfinement and chiral
symmetry restoration at $T_c$, nothing prevents the QCD coupling
from running to larger values at lower momentum scale until it is
stopped at the screening mass scale." The lattice scale is $\sim
(0.5\ {\rm fm})^{-1} \sim 400$ MeV because for higher scales
(shorter distances) the charges are locked into the quarks and
antiquarks. In order to compare with the lattice results on
$\alpha_s$ shown in Fig.~\ref{fig2}, we use the perturbative (long
distance) scaling
\be
\frac{\alpha_s (T\gsim T_c)}{\alpha_s
(T<T_c)} \simeq \frac{\ln (\Lambda{\chi}/\Lambda_{\rm QCD})} {\ln (400\
{\rm MeV}/\Lambda_{\rm QCD})} \simeq 3 \label{eqA1}
\ee
where we have
taken the chiral symmetry scale $\Lambda_\chi=1$ GeV
and $\Lambda_{\rm QCD} = 250$ MeV. The
actual increase in Fig.~\ref{fig2} from the $\alpha_s \sim 1/3$
for isolated charmonium to $\sim 8/3$ at $T_c$, where we have
included the Casimir operator of 4/3, is more than double this,
but at least of the right general size, leaving little doubt that
the increase in color Coulomb at $T_c$ results from movement
towards the infrared with chiral symmetry restoration. We show
below that the instanton molecule interaction lowers the scale
$\Lambda$ essentially to zero. BLRS\cite{BLRS} found that the
heavy quark Coulomb interaction provided only $\sim 1/8$ of the
attractive interaction which brought the $\pi$ and $\sigma$ masses
to zero at $T_c$, so the movement towards the infrared is much
more pronounced than given by the lattice results, which do not
include the velocity-velocity interaction and instanton molecule
interactions, neither of which would contribute appreciably to the
heavy quark situation.

The above discussion refers to the large distance behavior. The
shorter distance interaction is strongly influenced by the
necessity of the interaction going to zero as $r\rightarrow 0$,
because of asymptotic freedom.

Preliminary results on quenched QCD \cite{kaczmarek} and in
full QCD \cite{KEKLZ} are given in \cite{BNL}.
The chief result of full QCD was to rescale by the $T_c$ (unquenched);
i.e., results at $T_c (quenched) \sim 260$ MeV
were moved down to $T_c (unquenched) \sim 175$ MeV.
Whereas we believe this to be true for the heavy-quark color
Coulomb (singlet) interaction, we also believe that for other
problems the instanton molecule interaction may introduce
an additional scale, as we discuss below.

The lattice calculations show a great dependence upon radial
coordinate $r$, as might be expected from asymptotic freedom
and confinement. We showed the $\alpha_s$ from the Polyakov loops
in Fig.~\ref{fig2}, where $\alpha_s$ is extremely large at $T_c$.

However, when all is said and done, the singlet potential energy just
above $T_c$ can be schematized
\be
V=\frac{\alpha_s}{r}\;\;\; (r >\hbar /2m_q)
\label{eqA2}
\ee
with $\alpha_s=0.5$ to the accuracy at which we can read off the LGS
curves. This potential was the one used in BLRS, the Coulomb interaction
being modified for $r<\hbar/2 m_q$, which can be thought of as imposing
asymptotic freedom. Furthermore, this schematic potential should work
reasonably well up to $T\sim 1.4 T_c$, where the screening mass
scale can be read off as $\sim 0.5$ fm, distances below this scale
being important in the region of $T_c\sim 1.4 T_c$.
We believe that the above schematization is adequate to discuss physical
phenomena in this range of energies. From $T\sim 1.4-1.5 T_c$ the
Debye screening sets in, so that the instanton molecules become
bigger and by $T\sim 1.9-2 T_c$ they become unbound. We leave this
latter region to Shuryak and Zahed\cite{shuryak2003}.

Whereas we must increase the heavy-quark Coulomb interaction for
$T\gsim T_c$ by a factor of $\sim 8$ in order to include the
light quark effects, we have only the underpinning of the
heavy quark LGS, aside from our general argument that
$m_\pi= m_\sigma=0$ at $T=T_c$ in the chiral limit. The LGS show,
however, that the screening distance of $\sim 0.5$ fm is not
reached until $T\sim 1.4 T_c$, so that the charges are still
locked into the quark and antiquark up to this temperature to
a good approximation. This was all that was needed to construct
the instanton molecule interaction in BLRS, so we assume that
there is little change in the interactions up to this temperature,
although Debye screening sets in rapidly above.

In fact our argument that our scenario at $T_c$ should hold well
up to $1.4 T_c$ is improved if we construct the collective wave
functions. In BLRS we calculated the instanton molecule wave
functions with only the color Coulomb and velocity-velocity
potential; i.e., with an $\alpha_{s,\rm eff}=1$. This gave us an
rms radius for the $\pi$ and $\sigma$ of $\sim 1/3$ fm.
However, the energy of the $\pi$ and $\sigma$ were lowered only
$\sim 1/4$ of the way from 2 GeV to 1.5 GeV, as seen in Table~1
of BLRS.
We then calculated that the instanton molecule interaction would
further lower the energy to zero, but we did not calculate the
wave function of the collective state.

This wave function can be expressed in relative momenta $q$ of
quark and antiquark. If the energy is to be lowered by 2 GeV,
then this relative momenta must be made up of components that extend
to relative momenta of this size; i.e., the coherent wave
function, solution of the $\bar q q$-bubble sum, must be of radius
$\sim \hbar/2 m_q c$, or $\sim 0.1$ fm. In other words, the
collective wave function made up of all quark-antiquark,
or quark, quark-hole components must be really tiny if the collective
state it is to describe, is to be bound by $2 m_q c^2$.
Thus, the electric charge is really locked into a tiny volume, much
less at $T_c$ than the screening radius. This is why we believe that
there is little change in the $\pi$ and $\sigma$
masses between $T_c$ and $1.4 T_c$.

\begin{figure}
\centerline{\epsfig{file=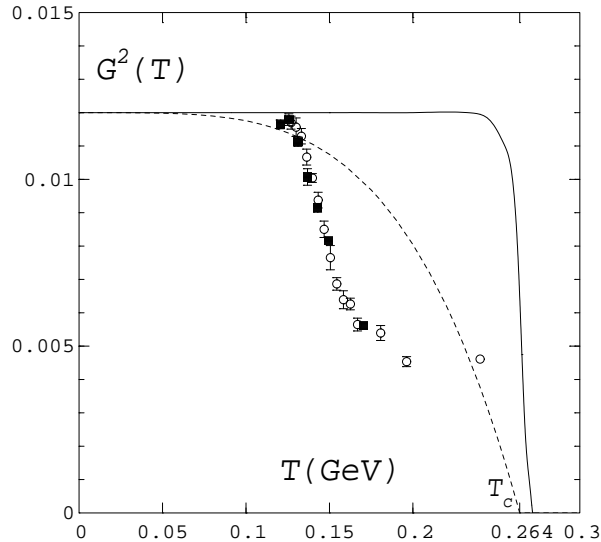,height=3in,
bbllx=85,bblly=240,bburx=484,bbury=611}}
\caption{Gluon condensates taken from Miller
(Fig.~2 of BGLR\protect\cite{BGLR}). The lines show the trace
anomaly for SU(3) denoted by the open circles and the heavier ones by
filled circles. The $T_c$ marked in the figure is that for
quenched QCD, whereas we deal with unquenched QCD with
$T_c=175$ MeV.}
\label{figA1}
\end{figure}

In fact, we see some indication of the constancy in gluon
condensate between $T_c=175$ MeV and $1.4 T_c$ from the LGS of
Miller\cite{Miller00} which we reproduce as our Fig.~\ref{figA1}.
The initial drop in gluon condensate (trace anomaly) up to
$T_c\simeq 175$ MeV is shown in BGLR\cite{BGLR} to be given by the
melting of the soft glue, the glue connected with chiral symmetry
restoration. It is connected quantitatively in BGLR with the
energy provided by the dynamical (or constituent) quark masses
going to zero. The last two points on the right show that the
``epoxy", the condensate of hard glue connected with instanton
molecules, does not change perceptively from $T_c=175$ MeV to $1.4
T_c$. This is just the region we discussed above in which the
electric charges are trapped in the small instanton molecules
inside of the screening radii. The large NJL forces, which bring
the molecule masses to zero or nearly to zero in the region $T_c$
to $1.4 T_c$ clamp the hard glue tightly in place.

We also got a simple estimate of the molecular breakup temperature
from our argument that quark and antiquark momenta $q$ must be
$\sim m_q c$. There will be break up once the thermal energy of $q$ is
equal to $m_q c^2$. But relativistically $q\sim 3 T$. Thus
$T_{\rm zb} \sim m_q c^2/3$, roughly $1.9 T_c$ (unquenched).

\begin{figure}
\centerline{\epsfig{file=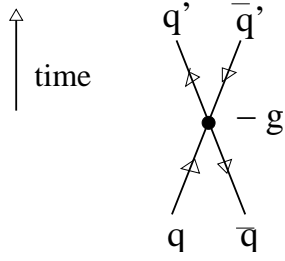,height=1.3in}}
\caption{Instanton molecule interaction $g$ which connects $\bar q q$
bound states.}
\label{figA2}
\end{figure}

We now repeat here our treatment of the instanton molecules and
their connection with the giant resonances found in the lattice
calculation of Asakawa et al.\cite{asakawa03} since
Ref.\cite{brown67} is out of print. We have established that our
$\bar q q$ (or quark, quark-hole) representation is that of the
Coulomb states. Every $\bar q q$ state is connected to every other
$\bar q q$ state by the 4-point instanton molecule interaction as
in Fig.~\ref{figA2}.

We construct this interaction either for quark-particle,
quark-hole states or for quark-antiquark states, as noted in BLRS.
Initially for simplicity we consider only states of total
momentum zero; i.e., the $\bar q q$ state has no translational
energy. The latter will be put in later.

As in BLRS we take the quark and antiquark masses to be 1~GeV,
considering all $\bar q q$ state energies initially at 2 GeV
to be lowered to 1.5 GeV by the color Coulomb interaction.
The diagonalized collective
state will turn out to be a boson state at zero energy and zero
momentum. We will then treat
this state as a boson, neglecting the incoherent states which
remain at 1.5 GeV, where their Boltzmann factors at $T_c$ are
so small that they can be neglected.

The instanton molecule interaction starts out as a nonlocal one,
the nonlocality being governed by the $\bar \psi\psi$ of the
instanton zero modes. The $\bar\psi\psi$ is sharply peaked,
mostly lying within a radius of
$r\sim \sqrt{2/5}\rho$, where $\rho\sim 1/3$ fm is the radius of the
instanton (Appendix of BLRS\cite{BLRS}). This regulates the nonlocality
and acts as a cutoff in the
possible momentum range of the $\bar q q$ states. The way
this is handled in BLRS is to use a $\delta$-function
4-point interaction, but to cut it down by a factor
$F= (0.75)^2$.
The underlying nonlocality restricts us to a band in momentum space.

We have, therefore, several approximations, but we are guided
in our numerics by the principle that the pion energy must
turn out to be zero in the chiral limit in order to make a
smooth chiral restoration transition. Of course, the
chirally restored $\sigma$-meson must also have zero mass
at $T=T_c$.

If we keep only $\bar q q$ states going forward in time
(ignore ground-state correlations) we have the quantum mechanical
problem to solve, for $T=T_c$,
\be
\left[ -0.5\ {\rm GeV}\ {\rm I}_{N\times N}- g_I\ \left(
\begin{array}{cccc}
1 & 1 & \cdots & 1 \\
1 & 1 & \cdots & 1 \\
\cdot & \cdot & \cdots & \cdot \\
1 & 1 & \cdots & 1
\end{array}
\right)_{N\times N}\right]\psi_{N} = E\psi_{N} .
\label{eqA3}
\ee
In other words, having reduced the instanton molecule interaction
to a 4-point one, all
matrix elements in the secular matrix are equal.
Here $g_I$ is the instanton molecule interaction.
The first
$-0.5$ GeV times the $N$-dimensional unit matrix ${\rm I}_{N\times N}$
is just the color Coulomb interaction.

Now from the principle of insufficient reason (all off-diagonal matrix
elements are equal) we know that one eigenfunction
(the collective one) is
\be
\psi_{\rm coll} =\frac{1}{\sqrt N}
\left(
\begin{array}{c}
1 \\
1 \\
\cdots \\
1
\end{array}
\right)_N
\label{eqA4}
\ee
and we easily find that
\be
E=-0.5\ {\rm GeV} - N\ g_I.
\label{eqA5}
\ee
In BLRS\cite{BLRS} we accomplished the same diagonalization by a
sum of $\bar q q$ loops going forward in time, giving
\be
N g_I = 0.75\ {\rm GeV}.
\ee
Thus, at this level the collective state is brought down from the
$2 m_q=2$ GeV to 0.75 MeV.

\begin{figure}
\centerline{\epsfig{file=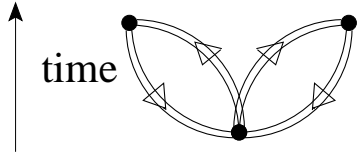,height=0.8in}}
\caption{Backward-going loops; equivalently, ground-state correlations.
}
\label{figA3}
\end{figure}

The equivalence between solving the secular matrix and summing loops
as was done in BLRS going
forward in time was developed in detail by Brown \cite{brown67}.
There it was shown that if the collective state is brought down
half-way to zero (In our above case from 1.5 GeV to 0.75 GeV
by forward going loops.) then with inclusion of backward going ones
it will be brought to $E=0$. The backward-going diagrams are
really ground-state correlations, are shown in Fig.~\ref{figA3}.

\begin{figure}
\centerline{\epsfig{file=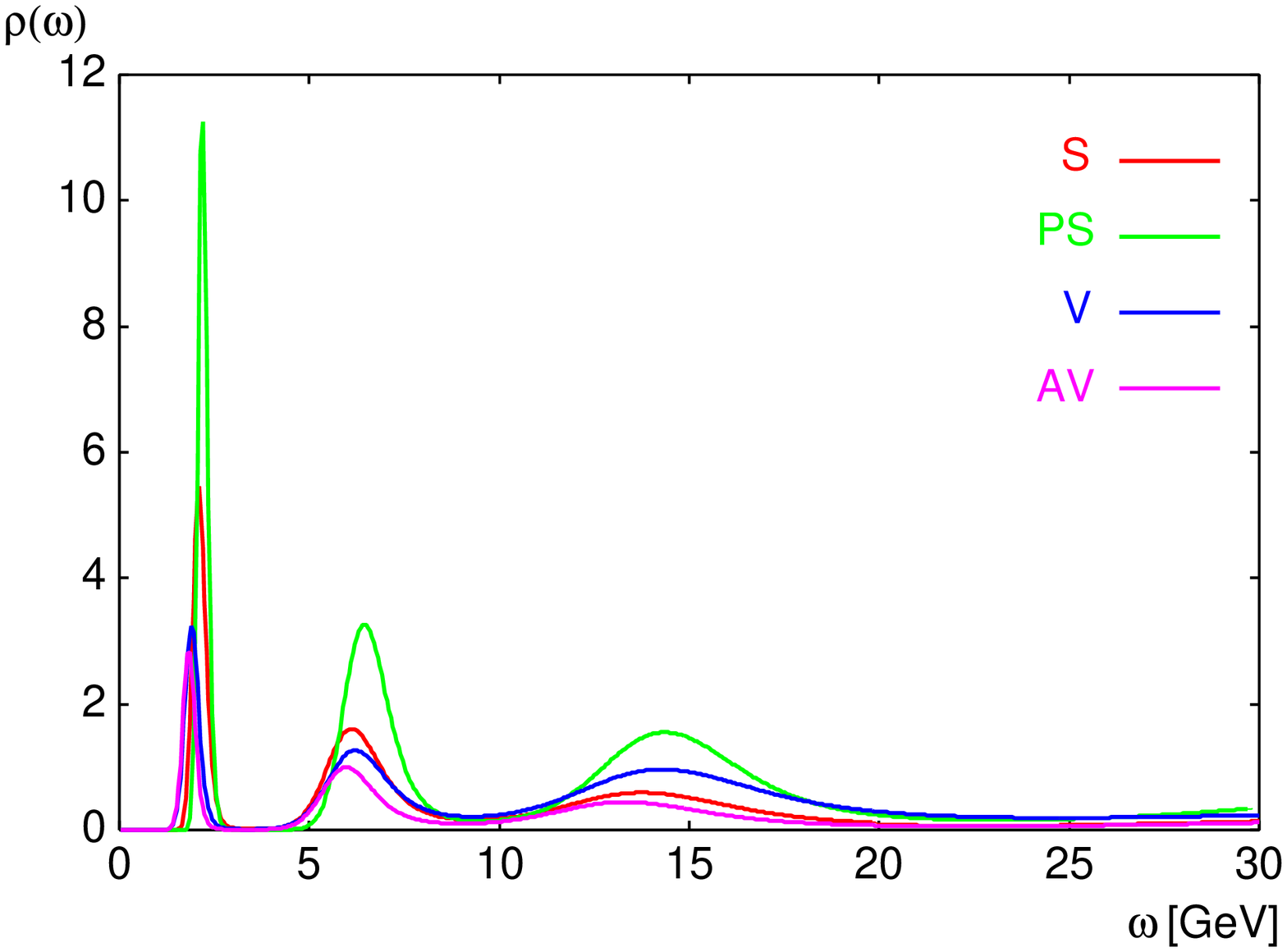,height=2in}
\epsfig{file=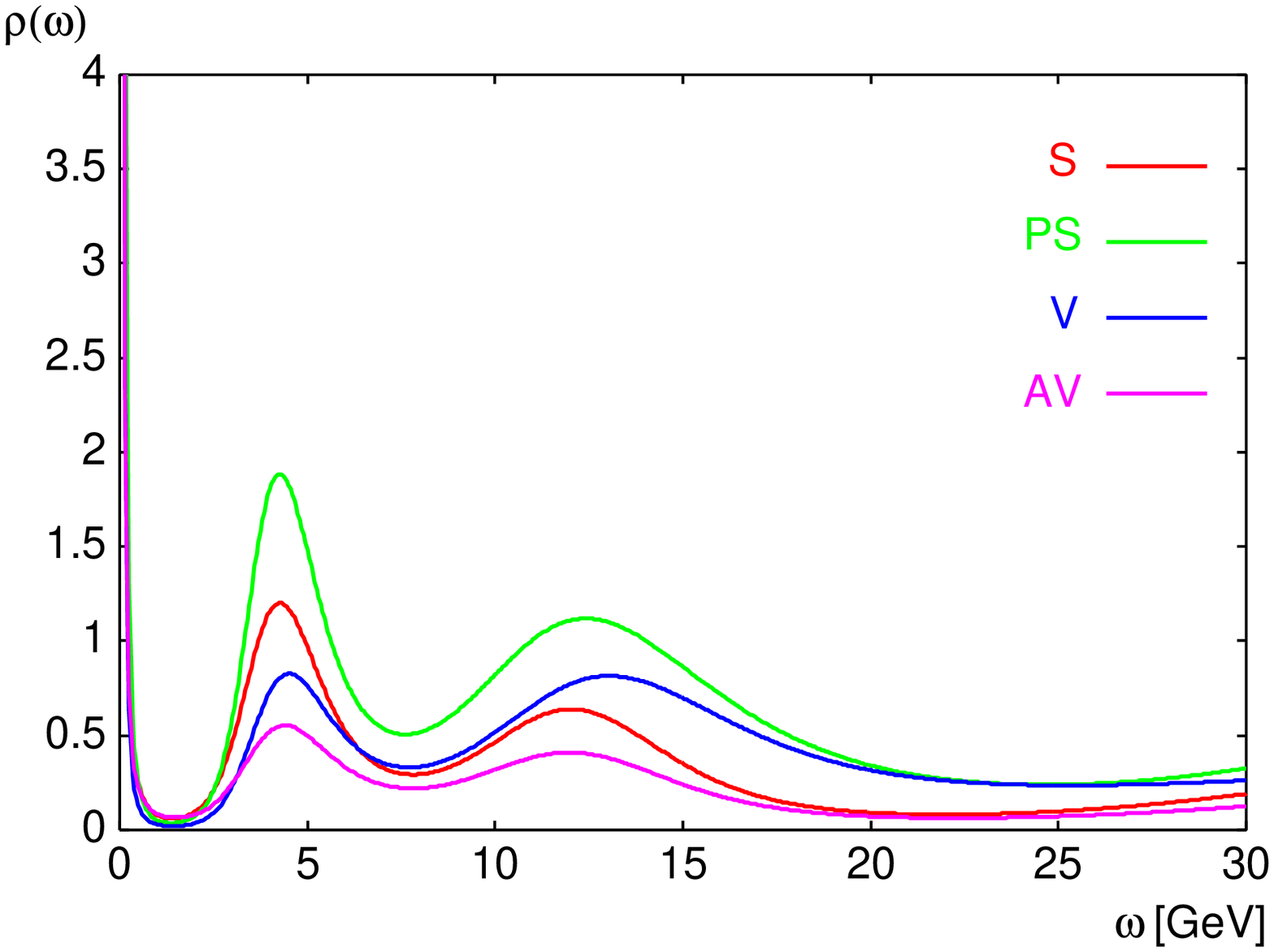,height=2in}}
\caption{Spectral functions of Asakawa et al.\cite{asakawa03}.
Left panel: for $N_{\tau}=54$ ($T\simeq 1.4 T_c$).
Right panel: for $N_{\tau}=40$ ($T\simeq 1.9 T_c$).}
\label{figA4}
\end{figure}

We consider only the fundamental mode of the vibrations, the
higher modes being lattice artifacts \cite{Datta03,CPPACS}.

Let us consider the left panel of Fig.~\ref{figA4}, taking the fundamental
to be at 2.1 GeV. As noted, BLRS\cite{BLRS} had no isospin dependence
and very small spin dependence, so the SU(4) character of the vibrations
is explainable. Now we developed above that the situation at
$T=1.4 T_c$ should not have changed much since $T_c$, because the
screening scale has only reached that of the instanton molecules
about there.

In fact, the calculation in BLRS\cite{BLRS} used the Coulomb wave functions
and then summed the loops with four-point interaction given by the
instanton molecule one. The very small collective wave function of
size $r\sim \hbar /2 m_q$ was not constructed. But the Coulomb
potential was chosen so that $dV/dr |_{r=0}=0$ and the
instanton
molecule source $\bar\psi\psi\sim (r^2+\rho^2)^{5/2}$ also had
$dV/dr=0$, both of which should follow from asymptotic freedom.
Furthermore, since the molecule energy is a minimum at $T_c$, it should
recover (from zero) only quadratically as $(T-T_c)^2$ from its minimum.
All of these effects make the growth in molecule slow as $T$ moves
above $T_c$, so that situation at $T=1.4 T_c$ should not be much
different from $T=0$.

Neglecting small changes from $T_c$ to $1.4 T_c$ in
$m_\sigma$ and $m_\pi$, we find that with $m_q=1.2$ GeV, giving unperturbed
molecule energy of $m_q+m_{\bar q} = 2.4$ GeV, then the heavy quark
Coulomb potential will bring this down $\sim 1/8$ of the way to zero,
or to 2.1 GeV. This $m_q$ is $\sim 20\%$ below the Petreczky et al. 1.5 GeV
mass from quenched calculation at 3/2 $T_c$, but still above the 1 GeV
that would be applicable if $m_q$ scaled with $T_c$ in going from
quenched to unquenched calculations. Although this difference is
probably not larger than uncertainties in the LGS, we believe that it
is indicative of another scale than the color Coulomb one, that of
the instanton molecules, between $T_c$ and $1.4 T_c$.

In any case, the fact that the LGS find a vibration of energy
$\hbar\omega \sim 2$ GeV at $T=1.4 T_c$ and that the vibration
requires attraction between quark and antiquark, shows that
$m_q> 1$ GeV.

Now using the fact that the LGS have only $\sim 1/8$ of the
total attraction, we find that the $m_\sigma$ and $m_\pi$ would
still be essentially zero at $T=1.4 T_c$. We confess that our above
scenario seems to fit a complicated situation too well, but
believe it to be roughly true.
Notice that at $T=1.4 T_c$, where we have argued the situation is not much
different from $T_c$ because the charges are locked into molecules inside
of the Debye screening, the LGS show a complete SU(4) symmetry.
Harada and Yamawaki\cite{HY:PR} show that in the chiral limit the
$\rho$ mass joins the $\pi$ and $\sigma$ mass as $T$ goes upwards
to $T_c$ at zero. It seems most reasonable that the $\rho$ mass
is also zero just above $T_c$.

Once the screening range of $\sim 0.5$ fm is reached by the size of the
instanton molecules, the latter would be expected to become rapidly
unimportant, probably by $T=1.9 T_c$ where the situation
appears to have gone perturbative \cite{BLRS} or at least to have
reached the breakup of the $\bar q q$ bound states. Since the
instanton molecules are unimportant at $1.9 T_c$, that temperature
should scale with $T_c$ in going from quenched to unquenched; thus
in full QCD it should be $1.9 T_c (unquenched) \simeq 332$ GeV,
just above the temperature at RHIC following the color glass stage.

In fact we referred to the LGS as having seen vibrations. From our
above discussion it should be clear that they saw only the
$\bar q q$ bound states, and only the binding of these by the Coulomb
interaction, not including the velocity-velocity interaction or
instanton molecule interaction which do not contribute for heavy quarks.
The real vibrations are those we calculated earlier in this section,
with inclusion of backward going graphs (ground-state correlations).
In order to get the full collectivity these latter effects have to be
included, but they cannot be in Euclidean time LGS. However, they can
be added theoretically as we did for $T=T_c$.

Thus, for the modes such as the $\pi$ and $\sigma$ which go massless
at $T_c$, the best that even an unquenched calculation (full QCD
for light quarks) can do is to bring them down halfway to zero
because the ground state correlations cannot be handled in Euclidean space.
(An exponentially decreasing antiquark cannot be a quark going
backward in time.)
(We noted above that in Minkowski space, if the forward going
graphs brought the vibrations down halfway, then the inclusion of
backward-going graphs would bring them down all the way.
Note that our schematic mode eqs.~(\ref{eqA3})-(\ref{eqA5})
is equivalent to a treatment with only forward-going graphs.)
In this case, because the diagonalization of the secular matrix
puts all of the trace into the one collective state for each
spin and isospin, the strength in this one collective state is equal
to the sum of the strengths in all of the individual $\bar q q$-molecules,
although the energy of the collective state will be moved down
4-times further at $T_c$.

The theory of vibrations has many names; e.g., random phase approximation,
particle-hole solutions of the Bethe-Salpeter equation, bosonization of
particle-hole excitations, time-dependent Hartree-Fock,
linearization of the equation of motion. The latter carries the best
description; they are the best that can be made from linearization.
They are also the lowest-order approximation that satisfies
conservation equations; e.g., conserves Newton's law\cite{baym62}.

It may seem somewhat amazing that we can apply the same theory of
vibrations in matter at temperatures of $\sim 200$ MeV as we do in
nuclei at low energies. However by eliminating longitudinal and
scalar gluons in favor of an instantaneous Coulomb interaction
\footnote{Note the movement towards the strong $N=4$
supersymmetric Yang-Mills theory at finite temperature in which
the scalar and longitudinal phases travel faster at superluminal
speeds as the coupling becomes stronger \cite{shuryak03b}.} which
gives the Furry interaction, and then, integrating over time and
regularizing by integrating over the spacelike nonlocality, one
gets an $\sim 50\%$ reduced, but instantaneous interaction
(Appendix of BLRS \cite{BLRS}). Thus, we have the same tools in
hand as in low-energy nuclear physics with the added simplicity
that our Furry representation quark-antiquark states are really
degenerate, as confirmed by the LGS. (In low-energy nuclear
physics the term energy in the giant dipole excitation
$\hbar\omega$ is the distance between shells but the unperturbed
energies are split by the spin-orbit interaction which gives added
structure\cite{brown49}.) Thus, we do have the same problem, but
heavy quark LGS have seen only a small part of it, as we laid out
above. Nonetheless, the small parts of the ``iceberg" seen in the
LGS give already the most prominent excitations above $T_c$.

There may, in fact, be an advantage in only a small part of the
light-quark interaction entering into the LGS. The vibrations look pretty
linear at $T=1.4 T_c$. However, with the $\sim 8$ times larger
interaction and the strong $\rho\leftrightarrows 2\pi$
mixing we have investigated, the widths of the $\rho$ and $\pi$
might have been spread out over such a wide range of energies that
the vibrations would not be discernible from the background.


\section{Appendix B: Consequences of Rescaling}
\label{appB}

We needed Appendix A in order to discuss the physics accompanying
rescaling, primarily to clarify that the LGS takes note of the
change in scale from $\Lambda_\chi =1$ GeV to the instanton
molecule (or, equivalently, chirally restored meson) scale set by
their zero masses, in the chiral limit. Thus, although we go up in
temperature to $T_c$, suddenly the scale $\Lambda_\chi$ governing
the behavior of our thermodynamic quantities drops out, and the
system reverts way back toward the infrared. However, until chiral
symmetry is restored, the system knows nothing about the infrared
behavior that enters at $T_c$.

Thus, within the chirally broken system there is a movement
towards weak coupling in the effective variables, i.e., the
hadrons, resembling asymptotic freedom. At the risk of a misnomer
(since non-asymptotic fixed point is involved), we shall call this
``effective-sector asymptotic freedom." Beta functions are written
down in these variables, the vector mesons replacing gluons, etc.
in the hidden local symmetry of Harada and Yamawaki\cite{HY:PR}.
The flows of various quantities in their renormalization group
treatment join smoothly with those of QCD, so their theory in the
effective sector mimics QCD, even to the point of having a local
gauge symmetry: The critical point $T_c$ delineates flavor gauge
theory from color gauge theory, the latter ceding to the former in
a smooth way~\footnote{This ``continuity" of gauge degrees of
freedom between color gauge and flavor gauge was already
conjectured in \cite{BR96}.}. One may say that the HLS theory is
the shadow of QCD in the effective sector. It even has its own
gluon condensate \cite{BGLR}, the soft glue which melts as the
temperature increases towards $T_c$ (see fig.~\ref{figA1}) although
the effective sector knows nothing of the hard glue, the
explicitly broken chiral symmetry above $T_c$. Thus, in their own
effective world the effective theory has a gauge theory
description, effective asymptotic freedom, etc. It gets its scale
from the spontaneous breaking of chiral symmetry through the
condensate of soft glue \cite{br91} as shown in Fig.~\ref{figA1}.

In fact, from the region in which $T\sim 125$ MeV up to $T_c=175$
MeV the system behaves as if it is in a mixed phase with nearly
constant pressure since as described in \cite{BGLR} the increase
in energy with temperature goes into melting the soft glue. Thus
the elliptic flow is very low and, indeed, collapses around
mid-rapidity for protons in the NA49 40 AGeV experiments
\cite{stocker04}; the velocity of sound $v=p/E$ decreases with
increasing energy. It is in this region that the Harada and
Yamawaki\cite{HY:PR} scenario of vector manifestation brings out
the effective sector asymptotic freedom. The hadrons are mainly
pions and vector mesons and the fixed point of $m_\rho=0$ and
$g_V=0$ as $T_c$ is reached from below comes in to further soften
the interactions. In BGLR \cite{BGLR} we showed that the nucleons
dissolved into constituent quarks as the soft glue is melted, the
quarks then going massless, in the chiral limit, as $T$ moves up
to $T_c$. As noted at several places in our paper, the pressure is
approximately, possibly exactly, zero at $T_c$. (The free energy
$F$ is essentially $-pV$ for our purpose, so making it zero makes
the pressure zero.) All of this is efficiently encoded in the HLS
with the vector manifestation, HLS/VM, i.e., the effective gauge
theory in effective variables. Whereas it is true that this theory
thus far describes only the behavior of pions and vector mesons,
and has not been extended to include baryons~\footnote{It has been
shown however that the VM holds also for the constituent quarks,
so the dynamical quark mass vanishes at the chiral
transition~\cite{HKR}.}, the former are the important variables
for the thermodynamics of the relativistic heavy ion collisions.

Now the effective sector given by the Harada-Yamawaki theory which
is not valid above $T_c$ knows nothing about the sector in which
chiral symmetry is restored. Before Harada and Yamawaki
RG\cite{HY:PR}, it was clear that something must change, because
the Hagedorn temperature limited the increase in $T_c$. What we
have shown is that simply to enforce the most basic (and simple)
requirement that the Goldstone boson, the pion, move up smoothly
through $T_c$ without changing mass, requires the weakly
interacting effective sector at $T_c-\epsilon$ to be replaced by
the most strongly interacting sector in the gamut of the system,
and this is what RHIC has done.

The ``new" system, representing bona-fide QCD, has been greatly
studied and described. It has its own gluon condensate, the hard
glue or epoxy, which explicitly breaks chiral symmetry. QCD has
its own running coupling constants, so the reincarnation of
asymptotic freedom, the shadow of which occurred just below $T_c$
in the behavior of the effective variables, occurs only higher up,
at $T\gsim 2 T_c$, possibly above the maximum temperature reached
at RHIC.

We thus see that our rescaling arguments reverse the generally
accepted scenario. Going up in temperature we first encounter a
weakly interacting region in the effective variables, just below
$T_c$. With chiral restoration this gives way to the most strongly
interacting matter\footnote{Recall that the pion mass is brought
down from $m_q+m_{\bar q}\sim 2$ GeV to zero by the interactions!}
encountered in heavy ion collisions.

With the reversal of the commonly accepted scenario, the weakly
interacting
system below $T_c$ changing into the strongly interacting system
just above $T_c$, it is clear that equilibrium has to take place in
the latter region. We have adduced arguments from LGS and from
experiment (especially the $\rho$-meson decay) to support this
new and surprising scenario.

\section{Appendix C: A Schematic Model for How the $\rho$ Goes On Shell}

Shuryak and Brown \cite{shuryakbrown} calculated that the
$\rho$-meson freezes out at a temperature of 120 MeV and a baryon
density of 15\% nuclear matter density. The $\rho$ mass at this
density was measured by STAR \cite{STAR} to be 700 MeV. The $\rho$
then decoupled from the system and got the remaining $\sim 10\%$
of its on-shell mass back from its kinetic energy. In the regime
below 125 MeV density dependent effects, which we do not discuss
here, dominate, but from 125 MeV up to $T_c=175$ MeV thermal
effects connected with the melting of the soft glue
predominate\cite{BGLR}. In this paper the constituent quark mass
was correlated with the binding energy of the soft glue (which is
responsible for the dynamical breaking of scale invariance; i.e.
Brown-Rho scaling. The soft glue was shown in Fig.~\ref{figA1}.
Shuryak and Brown (see Sec. II.E of Ref.~\cite{shuryakbrown})
argued that collision broadening built the $\Gamma^\star_\rho$
back up to the on-shell $\Gamma_\rho=150$ MeV at $T_{\rm freeze\;
out}=120$ MeV; the part of the width connected with the two-$\pi$
decay being brought down to 100 MeV because of the reduced
(p-wave) penetrability involved in the decay. We consider here
only the $\rho$ connection with the two-$\pi$ system.

In the Harada and Yamawaki\cite{HY:PR} scenario $m_\rho^\star$ and
$\Gamma_\rho^\star$ go to zero in the chiral limit as $T$ goes to
$T_c$ from below; also $g_V^\star/m_\rho^\star$ goes as a constant
as $T$ goes to $T_c$. This means that\footnote{
Note that from Sec.~II.E of Shuryak and Brown\cite{shuryakbrown},
$\Gamma_\rho^\star/\Gamma_\rho$ would go to zero as
$(m_\rho^\star/m_\rho)^3$ even if $g_V^\star$ did not scale,
because of the p-wave pion penetrabilities and reduced phase
space for decay. 
Indeed the expression for the decay of the free $\rho$ is
$$
\Gamma_\rho =\left(\frac{g^2}{4\pi}\right)
\frac{m_\rho}{12}\left(1-\frac{4 m_\pi^2}{m_\rho^2}\right)^{1/2}.
$$
If we replace $m_\rho$ by $m_\rho^\star$ and let the 700 MeV
nearly on-shell mass found at thermal freezeout by STAR\cite{STAR}
go to zero, this expression gives a $\Gamma_\rho^\star$
that scales rapidly dropping to 44 MeV by the time
$m_\rho^\star=350$ MeV. Thus, our decoupling of the $\rho$ from the
pions does not depend much upon the scaling of $g_V^\star$; matters
would not be much changed if one had only BR scaling.
}
\be
 \frac{\Gamma_\rho^\star}{\Gamma} \longrightarrow
\left(\frac{m_\rho^\star}{m_\rho}\right)^3
\left(\frac{g_V^\star}{g_V}\right)^2, \ee three powers
coming from the p-wave penetrability and two from $(g_V^\star)^2$.

We now make the assumption that the dynamically generated part of
the $\rho$-mass scales with the constituent quark mass in the
chiral limit. This is a reasonable assumption near $T_c$ as it was
near $n_c$~\cite{HKR}, but we extrapolate it all of the way until
the constituent quark gets nearly all of its mass back. {}From
BGLR\cite{BGLR} we have the binding energy of the soft glue going
as
 \be {\rm B.E.} = 12\int_0^\Lambda \frac{d^3k}{(2\pi)^3}
\left(\sqrt{k^2+{m_Q^\star}^2} - |k| \right).\label{C2} \ee
We recall that this equation is motivated by Nambu-Jona-Lasinio in
which the binding energy in the vacuum -- in this case that of the
soft glue -- is obtained in terms of the negative energy quarks in
the condensate which breaks chiral symmetry having dynamically
generated masses $m_Q^\star$.

For $m_Q^\star \ll k$ we get from (\ref{C2})
 \be {\rm B.E.}
\approx \frac{3}{\pi^2} \frac{\Lambda^2}{2} {m_Q^\star}^2. \ee

We then make the simplest schematic model incorporating the above
ideas as \be \Gamma_\rho^\star \approx 100\; {\rm MeV} \left[
\left(1 - \frac{2 m_\pi}{700\; {\rm MeV}}\right)
\left(\frac{T_c-T}{50\ {\rm MeV}}\right) + \frac{2 m_\pi}{700\
{\rm MeV}}\right]^5 \label{eqC4} \ee which comes to 100 MeV at
T=125 MeV rather than at 120 MeV as found by Shuryak and
Brown\cite{shuryakbrown}, but we want a point midway between 120
and 130 MeV to do our later integration as the STAR
experiment\cite{STAR} does not constrain $\Gamma$ accurately. (In
fact, there are indications that at densities $< n_0$ and probably
also at low temperatures -- see Fig.~2 of Brown and Rho
\cite{brownrho03} -- $g_\rho^\star$ does not scale as rapidly as
$m_\rho^\star$. This would move the point where
$\Gamma_\rho^\star=100$ MeV lower.) We then find the results in
Table~\ref{tabC1}, where $(\Gamma^\star)^{-1}$ gives an indication
of the mean free path in fermis.

\begin{table}
\begin{center}
\begin{tabular}{ccc}
\hline
T [MeV] & $\Gamma^\star$ [MeV] & $(\Gamma^\star)^{-1}$ [fm] \\
\hline
125  & 100 & 2.0 \\
135  & 53  & 3.8 \\
145  & 25  & 7.9 \\
155  & 11  & 18 \\
165  &  4  & 53 \\
\hline
\end{tabular}
\caption{Effective decay widths and mean free paths for various
temperatures.}
\label{tabC1}
\end{center}
\end{table}

As noted in BGLR\cite{BGLR}, this region of temperatures is in the
environment usually described as a mixed phase. Rather than being
at constant temperature $T_c=175$ MeV, where the bag constant is
usually assumed to be a source of energy going down in
temperature, so that it keeps the temperature constant in a real
mixed phase, in our case (as shown in the lattice calculations,
Fig.~\ref{figA1}) the energy is furnished over a relatively large
(175 - 125 MeV) range of temperatures,  giving an $\sim 50$ MeV
range of mixed phase. The usually assumed $\sim 5$ fm/c from phase
transition to freeze out, means that each interval of 10 MeV takes
the system $\sim 1$ fm/c to traverse. We see that the $\rho$
mesons are never really equilibrated, although their final decay
rate just before freezeout would be sufficient to equilibrate them
had it acted for the full $\sim 5$ fm/c. They mostly decay in the
last 2 fm/c, centered about $T= 130$ MeV; i.e., from 140 to 120
MeV, the latter the freezeout temperature. In that interval,
by taking the average mean free path as $2.5$ fm,
roughly $\exp(-2 {\rm fm}/2.5 {\rm fm})=0.45$ of the $\rho$ mesons
don't decay into two pions.
Now the pions are always in equilibrium (with themselves). Thus,
the rate at which pions create $\rho$ mesons in the last 2 fm/c
would be equal to the rate at which $\rho$ mesons go into pions,
were there equilibrium. But the pion equilibration doesn't depend
appreciably on the $\rho$ mesons, so we can use the equilibrium
$\rho_0/\pi^-=0.08$ \cite{rapp03}; i.e., $(1-\exp(-2/2.5))\times
(\rho_0/\pi^-) = 0.55 \times 0.08$ will be created from the pions
in the last 2 fm/c. (This reverse process uses the same
$\Gamma_\rho^\star$ as the $\rho$-decay.) Thus, at thermal
freezeout we have $0.45 (\rho_0/\pi^-)_{\rm initial}$ left, plus
the $0.55\times 0.08 (\rho_0/\pi^-)$ produced in the last 2 fm/c
from the pions, the sum of which we equate with the STAR result
\be 
0.45 \times (\rho_0/\pi^-)_{\rm initial} + 
0.55 \times 0.08 =
0.169\pm 0.037 \label{eqC5} \ee which gives \be
(\rho_0/\pi^-)_{\rm initial} =0.20-0.36. \label{eqC6} \ee The
systematic error of $\pm 0.037$ gives the rather large uncertainty
in the initial number of $\rho$'s. We come to the factor of $\sim
2-4$ increase over the Braum-Munzinger et al's 0.11 value of
$(\rho_0/\pi^-)$ equilibrated at 177 MeV assumed just below $T_c$.
In the text we used $m_\rho^\star = 2 m_\pi$ just above $T_c$ so
as to minimize the free energy, but with out estimated
$\Gamma_\rho^\star\sim 380$ MeV just above $T_c$ the
$m_\rho^\star$ will be spread over a broad peak lying somewhat
below the on-shell $\rho$ mass of 770 MeV.

We believe our schematic model illustrates the main physics:
\begin{enumerate}
\item [(i)] The explicit chiral symmetry breaking is important, giving
the $2 m_\pi$ in the $\rho$-mass which roughly gives the center
of the $\rho$ mass distribution.
\item [(ii)]
The interaction of the $\rho$ with the pions is negligible at
$T_c$, increasing as $T$ drops, and becoming large at the end
of the mixed phase as the $\rho$ is nearly on shell.
\item [(iii)] The initial abundance of $\rho$'s is large, because they
are equilibrated just above $T_c$ with a spread of masses well
below the mass of the on-shell $\rho$.
\end{enumerate}

Given (iii) it is very reasonable that some of this initial
abundance remains at thermal freezeout. In fact, we believe in the
lower limit of at least twice that of the standard scenario in
which the $\rho$ freezes out just below $T_c$, because with large
pion and $\rho$ chemical potentials it seems possible to nearly
preserve the factor 2 from chemical to thermal freezeout.

As suggested before, CERN, in finding a weakly interacting system,
was operating mainly in the mixed phase described in
Appendix~\ref{appB}. RHIC clearly comes down through the region of
temperatures well above $T_c$, through $1.4 T_c$ down to $T_c$
where we have our low-mass chirally restored mesons. In this $\sim
2$ fm/c the $\rho$-mesons have time to decay into dileptons. This
should substantially increase the invariant masses in our range up
to $\sim 400$ MeV and possibly higher.

We now go on to discuss how our scenario of equilibrium above
$T_c$ will affect the dileptons.

Vector dominance is known to be violated due to the vector
manifestation fixed point $a=1$ at finite density and/or
temperature \cite{HY:PR}. However the ``intrinsic" violation more
prominent in the baryonic sector~\cite{brown02} is not expected to
make a qualitative influence in the mesonic sector, so we will
simply adopt the vector dominance in our discussions. We need
however take into account that the vector operating on the
physical (finite density) vacuum
does not only create the vector meson (and $\gamma$-ray), but
there are other many-body vector excitations, especially the
Rapp/Wambach ``rhosobar." The latter is an $N^\star (1520)$
excitation coupled to a nucleon hole, with vector quantum numbers.
\begin{figure}
\centerline{\epsfig{file=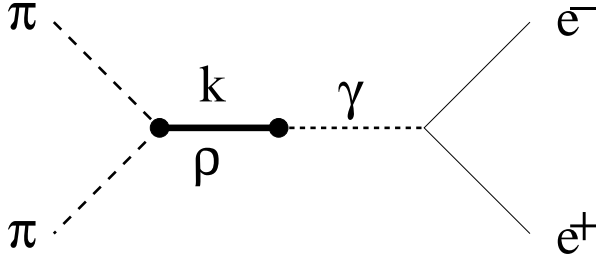,height=1.5in}}
\caption{Dilepton production through the $\rho$.}
\label{figC1}
\end{figure}
In fusing Brown/Rho and Rapp/Wambach \cite{brownrho03} Rapp
considered the production of rho mesons by pions, with the later
decay of the $\rho$ into dileptons in Fig.~\ref{figC1}. The
$\pi\pi\rho$ coupling involves a $g_V^\star$ and the
$\rho$-propagator, $(k^2+{m_\rho^\star}^2)^{-1}$. Thus, the vector
dominance coupling at the photon point $k^2=0$ is given by \be
g_{\rho\gamma}^\star =\frac{e {m_\rho^\star}^2}{g_V^\star}, \ee
for gauge invariance. Now as we go away from the photon point to
produce dileptons, in general the invariant mass of the latter is
$\sim m_\rho^\star$, the $\vec k$ of the photon being small
compared with the $k_0$. (Back-to-back kinematics involves only an
$\sim 10\%$ error in dilepton production.) In doing this, we
extend the vector dominance to the rhosobar, treated as a vector
particle.

\begin{figure}
\centerline{\epsfig{file=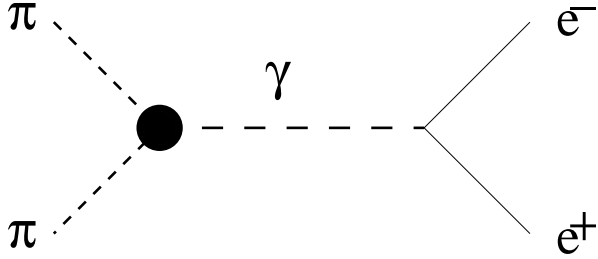,height=1.5in}} \caption{
Dilepton production via rhosobar dominance. The rhosobar
is subsumed in the big dot on the left.} \label{figC2}
\end{figure}

Thus, using vector dominance, Fig.~\ref{figC1} reduces to
Fig.~\ref{figC2}, where the $\rho$ being subsumed in the big dot
on the left.

We can check the validity of this approach at zero density and
temperature from the particle-data book where $\Gamma_\rho=150$ MeV
and the branching ratio into $e^+ e^-$ is
($4.49 \pm 0.22)\times 10^{-5}$, or
\be
\Gamma_{em}=6.7 \; {\rm keV}.
\ee
In Fig.~\ref{figC2} the branching ratio would be simply
$\alpha^2=5.3\times 10^{-5}$, a difference of only $\sim 10\%$.


We see from Tab.~\ref{tabC1} that $\Gamma_\rho^\star$ is large only
in the last part of the mixed phase, from $T=130$ to 120 MeV, the nearly
hadron phase as the $\rho$-meson goes back on shell.
We will call this the hadronic phase because the hadrons are nearly
on shell. We assume the $\Gamma_{em}=\alpha^2\Gamma^\star$,
so that dileptons come only from the hadronic phase in CERES.

Rapp and Wambach\cite{RW} enhance the dileptons substantially by
introducing the rhosobar, which at zero temperature puts $\sim 20\%$
of the $\rho$ strength at 590 MeV. With temperature and many-body
effects the width of the rhosobar broadens to $\Gamma\sim 250 $ MeV,
so that strength is moved down another $\sim 125$ MeV. With tails
of strength functions, a substantial number of dileptons are
produced down to $\sim 250$ MeV.

The factor ${m_\rho^\star}^2/g_V^\star$ must be introduced into
the amplitude for dilepton production if Brown-Rho scaling is to
be ``fused" with Rapp/Wambach. As noted by Brown and
Rho\cite{brownrho03} this factor cancels the enhancement that
would be obtained because of the B-R dropping $\rho$-mass, so that
Rapp/Wambach should fit well the hadronic dileptons.

We go into some detail with this because we believe that the hadronic
dileptons in RHIC will be similar to those from the 200 GeV/nucleon
CERES experiments. The scalar density in RHIC is not much less than
at CERN because the antibaryons give a large contribution in the former.
The factor of 2-4 in the $\rho$ abundance at $T_c$ will enhance
the RHIC dileptons. However, the chirally restored region of temperatures
will contribute at RHIC. In particular, the region of temperatures
from $T_c$ to $\sim 1.4 T_c$ will give dileptons in the range from
0 to $\sim 400$ MeV in our scenario, the broad region being given
by the large width \cite{asakawa03}. The contribution to the higher
dilepton invariant masses from higher temperatures will be smaller
because of the lower densities at the origin for these vibrations,
this density relevant for dilepton production.

\begin{figure}
\centerline{\epsfig{file=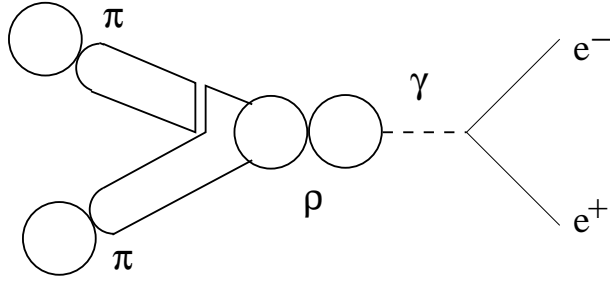,width=8cm}}
\caption{We redraw Fig.~\ref{figC1} in the chirally restored
region, $\pi+\pi\rightarrow \rho \rightarrow \gamma \rightarrow
e^+ e^-$, in the one-bubble addition we used to obtain the
$\Gamma=380$ MeV of eq.~(\ref{eq6}).} \label{figC3}
\end{figure}

The phenomenological vector dominance implies that a strong
interaction, in Fig.~\ref{figC2} the $\rho$-meson hidden in the
large black dot, precede a photon in the electromagnetic decay.
Since, as noted in BLRS\cite{BLRS} the vector mesons move smoothly
up through $T_c$, we believe that this assumption remains valid
above $T_c$ for the chirally restored mesons. In fact, the
chirally restored $\rho$ is the only vector particle above $T_c$,
the rhosobar having disappeared with the nucleons. In any case, it
seems the simplest way to couple dileptons to the $\rho$; namely,
to add an off-shell $\gamma$-ray to the $\rho$ (which is itself a
sum of bubbles in the BLRS random phase description) as in
Fig.~\ref{figC1}. This gives us the simple estimate for the
chirally restored sector of \be \Gamma_{em}\simeq \alpha^2
\Gamma_\rho \label{eqC9} \ee where we have estimated $\Gamma_\rho$
to be $\sim 380$ MeV at $T_c+\epsilon$ in eq.~(\ref{eq6}).

One can see that eq.~(\ref{eqC9}) is consistent with our calculation
giving eq.~(\ref{eq6}) of $\Gamma$ where we have added one bubble,
Fig.~\ref{fig5}, to each of the pions and $\rho$ in fig.~\ref{fig3}.
We redraw the final $\pi$'s and $\rho$ in Fig.~\ref{figC3},
to recover the analog of Fig.~\ref{figC1} in the chirally
restored region.

We thus expect about equal dilepton abundances from the chirally
restored and hardonic sectors, with very little contribution from the
mixed phase. The hadronic abundance should be about the same as in RHIC,
and that from the chirally restored sector to be at lower energies,
centered about $2 m_\pi$, but with a large width which causes the
upper end to overlap with the hadronic dileptons.

Because of the very strong coupling in the chirally restored
sector for $T\sim T_c - 1.4 T_c$, we cannot make any really
quantitative calculations, but we believe our schematic model to
give the main features. It will be exciting to confront them with
experiment.

Of course, we know that the dilepton ``cocktail" (background) is
more than an order of magnitude greater in the region about
dilepton invariant mass $\sim 2 m_\pi$, so the low mass dileptons
from the chirally restored sector may be difficult to separate
from these, but hopefully there will be distinguishing characteristics.



\begin{thebibliography}{99}

\bibitem{BMRS01} P. Braun-Munzinger, D. Magestro, K. Redlich,
and J. Stachel, Phys. Lett. {\bf B 518} (2001) 41.

\bibitem{HY:PR} M. Harada and K. Yamawaki, Phys. Rep. {\bf 381}
(2003) 1.

\bibitem{BLRS} G.E. Brown, C.-H. Lee, M. Rho and E.V. Shuryak, hep-ph/0312175;
Nucl. Phys. A, accepted.
{\it (denoted as BLRS)}

\bibitem{petreczky02} P. Petreczky, F. Karsch, E. Laermann, S.
Stickan and I. Wetzorke, Nucl. Phys. Proc. Suppl. {\bf 106} (2002)
513; hep-lat/0110111.

\bibitem{brown67} G.E. Brown, {\it "Unified Theory of Nucleon Models
and Forces"}, 1967, North Holland Pub. Co., Amsterdam.

\bibitem{asakawa03} M. Asakawa, T. Hatsuda and Y. Nakahara,
Nucl. Phys. {\bf A715} (2003) 863c.

\bibitem{wetzorke}
I. Wetzorke, F. Karsch, E. Laermann, P. Petreczky and S. Sickan,
Nucl. Phys. Proc. Suppl. {\bf 106} (2002) 510.

\bibitem{BMSW03} P. Braun-Munzinger, J. Stachel and  C. Wetterich,
nucl-th/0311005.

\bibitem{brown02} G.E. Brown and M. Rho, nucl-th/0206021;
Phys. Repts., in press.

\bibitem{vogl91} U. Vogel and W. Weise, {\it ``Progress in Particle and Nuclear
Physics"} {\bf 27} (1991) 195.

\bibitem{shuryak2003} E. Shuryak and I. Zahed, hep-ph/0307267.

\bibitem{Zantow} F. Zantow, 2004 University of Bielefeld Thesis, unpublished.

\bibitem{kaczmarek}
O. Kaczmarek, F. Karsch, P. Petreczky and F. Zantow, hep-lat/0309121.

\bibitem{BGLR} G.E. Brown, L. Grandchamp, C.-H. Lee and M. Rho,
Physics Reports, 391 (2004) 353. {\it (denoted as BGLR)}.

\bibitem{dewit} ``Field Theory in Particle Physics", Elsevier Science Publishers
1986, D. Dewit and J. Smith.

\bibitem{Kirson71}
M. Kirson, Ann. Phys. {\bf 66} (1971) 624.

\bibitem{brown49}
G.E. Brown and M. Bolsterli, Phys. Rev. Lett. 3 (1959) 472.

\bibitem{Bielefeld}
O. Kaczmarek, F. Karsch, P. Petreczky and F. Zantow, Phys. Lett. B
543 (2002) 41.

\bibitem{BJBP93} G.E. Brown, A.D. Jackson, H.A. Bethe and P.M. Pizzochero,
                Nucl. Phys. {\bf A560} (1993) 1035.

\bibitem{Redlich}
F. Karsch, K. Redlich, and A. Tawfik, Phys. Lett. {\bf B571} (2003) 67.

\bibitem{BMRS03}
P.B. Braun-Munzinger, K. Redlich, and J. Stachel, Invited review
for ``Quark Gluon Plasma 3", eds. R.C. Hwa and Xin-Nian Wang,
World Scientific; nucl-th/0304013.

\bibitem{STAR}
P. Fachini, for the STAR Collaboration, Nucl. Phys. A715 (2003) 462.

\bibitem{STAR2}
J. Adams et al., Phys. Rev. Lett. 92 (2004) 092301.

\bibitem{rapp03}
R. Rapp, Nucl. Phys. A725 (2003) 254.

\bibitem{Broniowski03}
W. Broniowski, W. Florkowski and B. Hiller, Phys. Rev. C68 (2003)
034911.

\bibitem{BSW91}
G.E. Brown, J. Stachel and G.M. Welke, Phys. Lett. B 253 (1991)
19.

\bibitem{br91} G.E. Brown and M. Rho, Phys. Rev. Lett. {\bf 66} (1991)
720.

\bibitem{BR03}
G.E. Brown and M. Rho, nucl-th/0305088; Phys. Repts., to be published.

\bibitem{isb}  K. Langfeld, H. Reinhardt and M. Rho, Nucl. Phys. {\bf A622} (1997)620.

\bibitem{Koch1993}
V. Koch and G.E. Brown, Nucl. Phys. A560 (1993) 345.

\bibitem{Karsch02}
F. Karsch, Nucl. Phys. A698 (2002) 199c.

\bibitem{KEKLZ}
O. Kaczmarek, S. Ejiri, F. Karsch, E. Laermann and F. Zantow,
hep-lat/0312015; Talk at Finite Density QCD, Nara, Japan, 10-12
July, 2003.

\bibitem{BNL}
Proc. of A RIKEN BNL Research Center Workshop, ``Lattice QCD at finite
temperature and density", Feb. 8-12, 2004, Brookhaven National Laboratory,
USA.

\bibitem{Miller00}
D.E. Miller, hep-ph/0008031; Phys. Repts. in preparation.

\bibitem{Datta03}
S. Datta, F. Karsch, P. Petreczky and I. Wetzorke,
hep-lat/0312037.

\bibitem{CPPACS}
T. Yamazaki et al. (CP-PACS Collaboration),  Phys. Rev. D65 (2002) 014501.

\bibitem{baym62} L.P. Kadanoff and G. Baym, ``Quantum Statistical
Mechanics", W.A. Benjamin, New York 1962.

\bibitem{shuryak03b}
E. Shuryak and I. Zahed, Phys. Rev. D69 (2004) 046005.

\bibitem{stocker04}
H. St\"ocker, private communication, to be published.

\bibitem{BR96} G.E. Brown and M. Rho, Phys. Rept. {\bf 269} (1996)
333.

\bibitem{HKR} M. Harada, Y. Kim and M. Rho, Phys. Rev. {\bf D66}
(2002) 016603.

\bibitem{shuryakbrown} E.V. Shuryak and G.E. Brown, Nucl. Phys. A717 (2003)
322.

\bibitem{brownrho03} G.E. Brown and M. Rho, Phys. Rept. 396 (2004) 1.


\bibitem{RW} R. Rapp and J. Wambach, Adv. Nucl. Phys. 25 (2000) 1.

\end{thebibliography}
\end{document}